\documentclass[fleqn,usenatbib]{mnras}

\usepackage{newtxtext,newtxmath}

\usepackage[T1]{fontenc}

\DeclareRobustCommand{\VAN}[3]{#2}
\let\VANthebibliography\thebibliography
\def\thebibliography{\DeclareRobustCommand{\VAN}[3]{##3}\VANthebibliography}

\usepackage{graphicx}	            
\usepackage{amsmath}	            
\usepackage{multirow}               
\usepackage{makecell}               
\usepackage{tikz,xcolor,hyperref}   

\definecolor{lime}{HTML}{A6CE39}
\DeclareRobustCommand{\orcidicon}{%
    \begin{tikzpicture}
    \draw[lime, fill=lime] (0,0)
    circle [radius=0.16]
    node[white] {{\fontfamily{qag}\selectfont \tiny ID}};
    \draw[white, fill=white] (-0.0625,0.095)
    circle [radius=0.007];
    \end{tikzpicture}
    \hspace{-2mm}
}

\newcommand{\invdays}{\textrm{d}^{-1}}
\newcommand{\starref}{${}^*$}
\newcommand{\daggerref}{${}^\dagger$}

\newcommand{\orcidRP}{\href{https://orcid.org/0000-0002-7057-0151}{\orcidicon}}
\newcommand{\orcidMC}{\href{https://orcid.org/0000-0001-6003-8877}{\orcidicon}}
\newcommand{\orcidKP}{\href{https://orcid.org/0000-0002-9372-5574}{\orcidicon}}
\newcommand{\orcidPP}{\href{https://orcid.org/0000-0002-8178-8463}{\orcidicon}}

\title[Machine Learning Methods for Variable Stars]{Semi-Supervised Classification and Clustering Analysis for Variable Stars}

\author[R. Pantoja et al.]{
R. Pantoja$^{1,3}$\thanks{E-mail: rdpantoja@uc.cl}\orcidRP,
M. Catelan$^{1,2,3}$\thanks{E-mail: mcatelan@astro.puc.cl}\orcidMC,
K. Pichara$^{4}$\thanks{E-mail: kpb@ing.puc.cl}\orcidKP,
and P. Protopapas$^{5}$\thanks{E-mail: pavlos@seas.harvard.edu}\orcidPP
\\
$^{1}$Instituto de Astrof{\'{\i}}sica, Facultad de F{\'{\i}}sica, Pontificia Universidad Cat\'olica de Chile, Av. Vicu\~na Mackenna 4860, 7820436 Macul,
Santiago, Chile\\
$^{2}$Centro de Astroingenier{\'{\i}}a, Pontificia Universidad Cat\'olica de Chile, Santiago, Chile\\
$^{3}$Millennium Institute of Astrophysics, Santiago, Chile\\
$^{4}$Departamento de Ciencias Computacionales, Facultad de Ingenier{\'{\i}}a, Pontificia Universidad Cat\'olica de Chile, Santiago, Chile\\
$^{5}$Institute of Applied Computational Science (IACS), Harvard University, Cambridge, MA, USA}

\date{Accepted XXX. Received YYY; in original form ZZZ}

\pubyear{2022}

\begin{document}
\label{firstpage}
\pagerange{\pageref{firstpage}--\pageref{lastpage}}
\maketitle

\begin{abstract}
The immense amount of time series data produced by astronomical surveys has called for the use of machine learning algorithms to discover and classify several million celestial sources. In the case of variable stars, supervised learning approaches have become commonplace. However, this needs a considerable collection of expert-labeled light curves to achieve adequate performance, which is costly to construct. To solve this problem, we introduce two approaches. First, a semi-supervised hierarchical method, which requires substantially less trained data than supervised methods. Second, a clustering analysis procedure that finds groups that may correspond to classes or sub-classes of variable stars. Both methods are primarily supported by dimensionality reduction of the data for visualization and to avoid the curse of dimensionality. We tested our methods with catalogs collected from OGLE, CSS, and Gaia surveys. The semi-supervised method reaches a performance of around 90\% for all of our three selected catalogs of variable stars using only $5\%$ of the data in the training. This method is suitable for classifying the main classes of variable stars when there is only a small amount of training data. Our clustering analysis confirms that most of the clusters found have a purity over 90\% with respect to classes and 80\% with respect to sub-classes, suggesting that this type of analysis can be used in large-scale variability surveys as an initial step to identify which classes or sub-classes of variable stars are present in the data and/or to build training sets, among many other possible applications.
\end{abstract}




\section{Introduction}\label{sec:intro}

Variable stars are fundamental tools in astrophysics which can provide us with essential physical properties of stars such as distance (hence luminosities), mass, radius, temperature, and evolutionary state. All of this unquestionably help us improve our present understanding of stellar evolution theory, the distance scale, and Milky Way/Local Group spatial structure. The discovery and study of variable stars have been thorough since the last century, setting the current stellar pulsation theory and both helio- and asteroseismology \citep[e.g.,][]{shapley1914,eddington1918,cox1980,CD-2002}. Thus,  variable stars provide a powerful method to probe stellar interiors \citep[see, e.g.,][for extensive recent reviews and references]{aerts2010,aerts-2021,catelan2015,CD-2021}.

The search for these stars and other transient objects in the sky has motivated numerous large-scale surveys that have had a tremendous impact on astrophysics in the last several decades. This includes, for instance, among many others, the Optical  Gravitational Lensing  Experiment \citep[OGLE;][]{ogle3}, the Massive Compact Halo Objects (MACHO) survey \citep{macho}, the All-Sky Automated Survey \citep{asas};  the Catalina Sky Surveys \citep[CSS,][]{larson2003} and the associated Catalina Real-time Transient Survey \citep[CRTS,][]{drake2009}, the Super Wide-Angle Search for Planets \citep[SuperWASP,][]{superwasp},  the Wide-Field Infrared Survey Explorer \citep[WISE,][]{wisemission}, the {\em Gaia} survey \citep{gaiamission}, and the VISTA Variables in the Via Lactea (VVV) survey \citep{vvvsurvey}. All of these surveys have led to a significant increase in the rate of discovery of new variable stars,  even though for some of those surveys, their original goals were to search for microlensing events, asteroids, near-Earth objects, and/or extrasolar planets. The immense data flow from these surveys is just the beginning, and it will be even more evident with the Vera C. Rubin Observatory's Legacy Survey of Space and Time (LSST,~\citealt{lsst, lsst-2019}), which will accumulate an enormous 30 TB of data per night for about ten years. Accordingly, a manual inspection of  million light curves to be obtained is an impractical task, so it is urgent to develop reliable, fast, and scalable algorithms to find and classify variable stars automatically.

Supervised learning classification algorithms are the most common among recent approaches used for variable star classification \citep[e.g.,][]{brink2013,hassan2013,pichara2013,xu2013,kuminski2014,elorrieta2016,pichara2016,benavente2017,castro2018}. In this context, the random forest algorithm achieves excellent classification accuracy \citep[i.e., $>90\%$;][]{debosscher2007, richards2011,dubath2011, kim2016,jayasinghe2019} when several expert human labeled variable stars are present, namely the training set. The main weak points of these methods are the following. On the one hand, the building of a training set can be expensive, time-consuming, and biased by human error, and as a result includes noise from other unknown classes. On the other hand, the design of suitable features to describe the light curves can be a challenging task. However, until recent years, the latter is gradually being resolved with the development of new neural network architectures for variable stars, demonstrating their capacity to learn useful features from the light curves and perform classification \citep{mackenzie2016,aguirre2019,becker2020}.

In contrast, unsupervised learning does not need labeled information to gain insights from data. This task is usually performed to assess the data structure, answer specific questions, or find a more compact representation of the data. There are three main types of unsupervised learning: dimensionality reduction (DR), manifold learning, and clustering. With clustering analysis, we can identify intricate patterns in the data without any prior knowledge or training set, partitioning it into groups or ``clusters'' that share such commonalities, which can be later analyzed and explored. Therefore, this task is ideal for data exploration, knowledge discovery, and outlier detection \citep{saxena2017}. Its main weakness is that we need some assumptions to explore the data first (e.g., the number of clusters) and well-designed features that are suitable for the final goal (e.g., to find and identify specific classes or outliers). In addition, evaluation of clustering analysis results is not trivial due to the lack of a training set. However, many of these flaws can be alleviated by modern non-linear DR algorithms and manifold learning. As an example, Uniform Manifold Approximation and Projection (\textsc{umap},~\citealt{umap}) provides unsupervised methods to reduce the data's dimensionality down to any desired number (e.g., two or three, to facilitate data visualization). Reducing data dimensions can be critical to evade the curse of dimensionality when dealing with many dimensions or features \citep{jimenez1998}, where points in Euclidean space progressively become uniformly distant from each other, making this metric inadequate for measuring dissimilarity between data points.

In the field of variable stars, unsupervised learning has been used for feature learning of light curves \citep{mackenzie2016} and for querying variable stars ranked by similarity \citep{valenzuela2018}. Recently, \cite{webb2020} presented an unsupervised anomaly detection method to detect transient events. For this purpose, they used the isolation forest algorithm \citep{liu2012} to find these sources, and the hierarchical density-based spatial clustering of applications with noise~(\textsc{hdbscan},~\citealt{hdbscan}) algorithm. At present, there is still significant untapped potential in clustering analyses applied to data acquired in the course of large-scale variability surveys. In fact, clustering could be insightful beyond variable star classification; other possible applications include the search for anomalous or transient objects, the understanding of light curve shapes, fast building of high-quality training sets, testing the current variable star taxonomy, discovering new classes or sub-classes of variable objects, among many others.

Semi-supervised learning algorithms broadly explore the range of possibilities in between the supervised and unsupervised approaches. This approach is used to solve classification tasks when there is a large amount of unlabeled data, and the labeled data are scarce \citep{chapelle2006}. There are just a few examples of the use of semi-supervised learning for variable star classification. This includes \cite{rimoldini2019}, where a semi-supervised approach is used to classify variable stars in the {\em Gaia} Data Release 2 (GDR2;~\citealt{gaiadr2}), and \cite{hoffman2019}, which implements a semi-supervised latent variable model for the classification of variable stars, achieving very high accuracy.

One of the main issues concerning the classification of variable stars is that a well-sampled and balanced training set is not always available. For instance, some classes are intrinsically more numerous than others, and variable stars with very short or very long periods may be very difficult to properly detect and characterize, depending on the cadence of each survey. A further source of error is introduced by partially finding variable stars through cross-matching with external, heterogeneous databases, such as provided by the General Catalog of Variable Stars \citep{samus2017}, the Variable Star Index \citep{watson2006}, and others \citep[e.g.,][]{marrese2019}. Indeed, this imbalance problem in classification has been approached recently by several authors using algorithmic or data augmentation techniques to improve the classification accuracy \citep[e.g.,][]{hosenie2019,hosenie2020}. On the other hand, the difficulty involved in gathering a well-sampled and unbiased training set has compelled other researchers to carry out classification of variables by visual inspection. This task is getting increasingly more infeasible as new surveys become ever more efficient in the acquisition of data for large numbers of previously unknown variable objects.

This situation is perhaps especially evident in the near-infrared, a wavelength regime which only recently, with the onset of telescopes such as VISTA and WISE, has started to be the subject of wide-field variability surveys \citep{vvvsurvey,Cioni-2011}. The near-infrared regime provides accordingly a suitable scenario in which to apply semi-supervised or clustering approaches, as the number of stars with suitable light curves available for training still remains insufficient for traditional supervised methods \citep[][and references therein; but see also \citealt{molnar2022}]{angeloni2014}.

In this work, we present a clustering analysis and semi-supervised classification method for variable stars, using data from three selected catalogs. With our clustering analysis, we assess the feasibility of finding pure clusters of classes or sub-classes of variable stars and other outlier objects. With our semi-supervised approach, we aim to evaluate the effectiveness of classifying large groups of variable stars based on small to medium-size training sets. The paper is divided into seven main sections. Section~\ref{sec:data} describes our datasets, the preprocessing, and the training sets used in our experiments. In Section~\ref{sec:features}, several features are extracted from the light curves, including a new set of features based on the periodogram. Section~\ref{sec:viz} presents visualizations of our datasets using the \textsc{umap} algorithm.
In Section~\ref{sec:clustering}, we present a new procedure to carry out our clustering analysis and the results on our datasets. In Section~\ref{sec:semisupervised}, we describe our semi-supervised clustering method and the results of its application to our adopted datasets. Finally, Section~\ref{sec:discussion} discusses our main findings and possible paths for future improvement. 

\section{The Data}\label{sec:data}

We selected catalogs of variable stars of very distinctive large-scale surveys to test and validate our semi-supervised classification and clustering analysis. Therefore, we chose three catalogs of variable stars of different sizes, cadences, number of observations per star, and passbands. This will guarantee diverse testing scenarios to gain a broad understanding of our methods. This section will briefly describe relevant aspects of each selected survey and its corresponding catalog of variable stars. The reader is referred to Appendix~\ref{appendix:nomeclature} for the different variable star classes that are included in this work, along with the adopted nomenclature and acronyms. 

\subsection{Catalina Sky Surveys}

Our first choice is the CSS, a survey with the primary goal of discovering near-Earth objects and potential hazardous asteroids \citep{larson2003}. The observations were carried out in a broad range of $-75^{\circ} < \delta < 70^{\circ}$ and $|b| \gtrsim 15^{\circ}$ in the sky for more than seven years. They were done continuously, stacking up time-series photometry only in the $V$ band (V\_CSS) leading to the finding of over 5  million variable stars candidates \citep{drake2014}. 

We compiled the main public catalogs of variable stars resulting from this survey, as obtained by the CRTS team \citep{drake2009}. These include the CSS periodic variable star catalog \citep{drake2014} and the CSS southern periodic variable star catalog \citep{drake2017}. We refer to these collective as the CSS catalog of variable stars (CSSCVS). This catalog comprises six classes of variable stars, subdivided into 16 sub-classes, with an average of 210 observations per star. There are many stars that were not assigned a class or a sub-class in CSSCVS, which could be problematic when training or evaluating the performance of our methods. Accordingly, these sources were not included in our analysis. 

\subsection{{\em Gaia}}

Our second choice is the {\em Gaia} survey \citep{gaiamission}, a European Space Agency astrophysical space mission that aims to provide accurate positions, parallaxes, photometry, and proper motions for more than a billion sources in our galaxy and beyond. It also collects spectroscopic data for millions of these stars. Its main objective is to build a 3D map of the Milky Way, in order to acquire an unprecedented understanding of its formation, dynamics, and evolution. Photometric observations cover the entire sky in the $G$-band, and low-resolution spectro-photometry is obtained for nearly all sources with blue and red photometers ($BP$ and $RP$, respectively).

GDR2 has on average 30 photometric measurements taken over 22 months for around $10^9$ sources \citep{gaiadr2}. The small number of observations per star limits the periodogram usage for in-depth asteroseismological studies, but this is compensated by the lack of atmospheric seeing and period aliases that are typically found in ground-based time-series observations. In our work, we adopted the variable star classification provided by \cite{rimoldini2019}, whose catalog (henceforth GDR2CVS) contains four classes and nine sub-classes of variable stars, with a median of about 23 observations per star. Included in our work are all GDR2CVS light curves with twelve or more observations in the $G$-band. 

\subsection{Optical Gravitational Lensing Experiment}

Our third choice is the OGLE Data Release III \citep{ogle3}, a long-term, large-scale sky survey firstly focused on searching for gravitational lenses, microlensing events, and variable stars. It quickly expanded to other fields, such as extrasolar planets, transient objects, structure of the Galaxy and Magellanic system, active galactic nuclei, interestellar extinction, Kuiper belt objects, and astrometry, among others. The observations in the $V$ and $I$ bands were made in 4 regions, the Milky Way's bulge, small fields in the Milky Way's disk, and the Magellanic Clouds. The OGLE Data Release III was completed in about twenty years, with different cadence configurations for each region in many seasons. The $I$-band's short cadence allows probing frequencies even over $20\ {\rm d}^{-1}$, but with a diverse quality due to the heterogeneity in the number of observations per star, which range between $\sim$20 to $\sim$3000, averaging 780. On the other hand, The $V$-band observations vary from $\sim$5 to $\sim$350, averaging 33.

One of the main results on variable stars from this survey is the OGLE-III collection of variable stars (OCVS,~\citealt{sozy2015}). OCVS also includes eclipsing binary star catalogs that cover the disk of the Milky Way \citep{pietru2013}, the Small Magellanic Cloud (SMC,~\citealt{pawlak2013}), and the Large Magellanic Cloud (LMC,~\citealt{graczyk2011}). The OCVS contains ten main classes of variable stars, which have more than 20 sub-classes in total. We rejected light curves with less than twelve observations in the $I$-band and two observations in the $V$-band. We did not include R~Coronae Borealis or $\alpha$~Canum Venaticorum variables because they contain no more than 25 exemplars in OCVS.

\subsection{Preprocessing} 

A preprocessing was performed to the light curves to remove unphysical quantities, outliers and fix other inconsistencies. The following measures were adopted to preprocess light curves for further feature extraction:
\begin{itemize}
    \item \textbf{Rejecting extreme values:} Magnitudes, times, or uncertainties with unphysical values were removed. In particular, we identified and excluded some stars with magnitudes and photometric errors beyond their corresponding survey’s limits, and epochs registered before the start date of the survey.
    \item \textbf{Removing duplicated observations:} Some light curves presented repeated observations. In such cases, we keep only the first observation of the original sequence.    
    \item  \textbf{Sorting:} Light curves are sorted by time. This is crucial for features that assume an ordered sequence.
    \item  \textbf{Removing outliers:} At most, three extreme observations for OCVS and CSSCVS and one for GDR2CVS were rejected from the light curve if they were over/below the median magnitude plus/minus two times its interquartile range (IQR).
\end{itemize}

\subsection{Final sets}

The sub-classes used in this work are constructed rearranging those provided by the catalogs into relatively larger groups for ease of visualization and training set assembling. Then, we build training sets drawing a stratified random sample of $5\%$ of each catalog's size. Finally, we under-sampled the majority of the sub-classes by limiting them to a maximum of 1500 stars each. In this way, we are setting a realistic upper limit for the number of stars that can usually be built via cross-matching with small to medium-size catalogs.

The general properties of the catalogs and training sets are shown in Table~\ref{tab:catalogs}, giving the catalog's number of stars per class and sub-class, and the corresponding number of stars used in the training sets.

\begin{table*}
    \caption{Catalogs properties and training set sizes}
    \label{tab:catalogs}
    \begin{tabular}{ccrrr}
    \hline\hline
    \textbf{Class} & \textbf{Class Total} & \textbf{Sub-Class} & \textbf{Sub-Class Total} & \textbf{Training Set} ($\lesssim 5\%$) \\
    \hline\hline
    \multicolumn{5}{c}{\textbf{CSSCVS}} \\
    \hline
    \multirow{3}{*}{ECL}   & \multirow{3}{*}{59080}  & EW/EB    & 49675    & 1500             \\
                                                      &                         & EA       & 9320          & 466              \\
                                                      &                         & PCEB     & 85            & 4              \\ \hline
    \multirow{3}{*}{RRLYR} & \multirow{3}{*}{16932}  & RRc      & 9185          & 459             \\
                                                      &                         & RRab     & 6745          & 337              \\
                                                      &                         & RRd      & 1002          & 50              \\ \hline
    LPV                    & 1798                    & \dotso  & 1798          & 90             \\ \hline
    \multirow{2}{*}{ROT}   & \multirow{2}{*}{1656}   & RSCVn    & 1514          & 76             \\
                                                      &                         & ELL      & 142           & 7              \\ \hline
    \multirow{3}{*}{CEP}   & \multirow{3}{*}{502}    & T2       & 277           & 14             \\
                                                      &                         & A        & 215           & 11              \\
                                                      &                         & T1       & 10            & \dotso              \\ \hline
    DSCT                   & 396                     & \dotso  & 396           & 20             \\   
    \hline    \multicolumn{5}{c}{\textbf{GDR2CVS}} \\
    \hline
    \multirow{3}{*}{RRLYR} & \multirow{3}{*}{177690} & RRab      & 144834    & 1500              \\
                                                      &                         & RRc      & 31929      & 1500               \\
                                                      &                         & RRd      & 927        & 47              \\ \hline
    Mira/SRV               & 149257                  & \dotso  & 149257     & 1500              \\ \hline
    \multirow{3}{*}{CEP}   & \multirow{3}{*}{8509}   & T1        & 6476      & 324               \\
                                                      &                         & T2    & 1721          & 86                \\
                                                      &                         & A      & 312          & 16               \\ \hline
    DSCT/SXPHE             & 8236                  & \dotso  & 8236         & 412           \\                           
    \hline
    \multicolumn{5}{c}{\textbf{OCVS}} \\
    \hline                          
    \multirow{3}{*}{LPV} & \multirow{3}{*}{330783} & OSARG     & 281387          & 1500             \\
                                                  &                         & SRV       & 42967           & 1500             \\
                                                  &                         & Mira      & 6429            & 321              \\ \hline
    \multirow{4}{*}{RRLYR} & \multirow{4}{*}{42761}  & RRab      & 30250           & 1500             \\
                                                  &                         & RRc       & 9825            & 491              \\
                                                  &                         & RRd       & 1319            & 66               \\
                                                  &                         & RRe       & 1367            & 68               \\ \hline
    \multirow{3}{*}{ECL}   & \multirow{3}{*}{38288}  & ED        & 23456           & 1173             \\
                                                  &                         & EC        & 8384            & 419              \\
                                                  &                         & ESD       & 6448            & 322              \\ \hline
    \multirow{6}{*}{CEP}   & \multirow{6}{*}{8645}   & T1$_{\rm F}$    & 4439            & 222              \\
                                                  &                         & T1$_{\rm 1O}$ & 2871            & 144              \\
                                                  &                         & T2        & 592             & 30               \\
                                                  &                         & T1$_{\rm M}$  & 564             & 28               \\
                                                  &                         & T1$_{\rm 2O}$ & 97              & 5                \\
                                                  &                         & A         & 82              & 4                \\ \hline
    \multirow{2}{*}{DSCT}  & \multirow{2}{*}{2808}   & S         & 2675            & 134              \\
                                                  &                         & M         & 133             & 7                \\ \hline
    DPV                    & 136                     & \dotso   & 136             & 7      \\ \hline
\multicolumn{5}{l}{\footnotesize \textbf{Notes}: For details about the class nomenclature, refer to Appendix~\ref{appendix:nomeclature}.}
\end{tabular}
\end{table*}

\section{Feature Extraction}\label{sec:features}

The engineering of light curve-based features is necessary to represent them as vectors of the same length to feed machine learning algorithms. The unevenly sampled nature of these data due to observation constraints and the presence of extended gaps between the main observation seasons makes this a challenging problem \citep{castro2018}. Conventional time-series analysis methods have to be adapted to this context in order to function correctly. Furthermore, in clustering, the design of features can be more complicated since cluster fragmentation, undesirable merging, and/or unexpected clusters could arise with the inclusion of features proven to work in supervised contexts. These undesired properties of the data cannot be ignored; thus, we can only return to feature selection and engineering to try to minimize this source of noise~\citep{aggarwal2013}. In this section, we describe our efforts to select the features available from different sources and the way we adapt them to match our prime goals. Moreover, we introduce a new approach to extract useful features from the light curve's periodogram.

\subsection{Light curve features}

There are several features for time series analysis readily available in the literature, e.g. Feature Analysis for Time Series \citep[FATS;][]{nun2015}, Abbe value features \citep{mowlavi2014}, the Cesium library \citep{naul2016}, even statistics features \citep{ferreiralopes2017}, Fourier parameters \citep{debosscher2007}, principal components analysis modelling features \citep{deb2009}, among others. Many of these features are proven to be excellent for supervised variable star classification. Although all of the above features were tested for our method, most were discarded since they induced severe cluster fragmentation or were biased by containing cadence information. Thus, we arrived at a stable set of features, but some cadence-correlated clusters may persist when the range in the number of observations per light curve is wide. This occurs because many features change their expected statistical properties depending on the number of observations in the light curve.

For clustering, it is necessary to transform some features that have extremely skewed values. Taking the logarithm of these features causes the range of their values to be reduced; as a result, they will have similar weight in the pairwise distance calculation computed by the machine learning algorithms used in this work. In other words, feature scaling is a way to balance the feature's relative importance. In Table~\ref{tab:features}, we briefly describe each set of features selected for our work and its corresponding reference.

Additionally, we formulated a feature that works reasonably well to distinguish some ECL from other types of variables, called the upper outlier fraction. It is simply the fraction of points over the third quartile plus 1.5 times its IQR. This feature helps to detect ECL variables that have a skewed distribution in magnitude. The observations that occur during narrow eclipses can be viewed as outliers in these distributions. It is worth noticing that the light curve features in Table~\ref{tab:features} were used for all the catalogs except in the case of CSSCVS, for which color-based features could not be used as CSSCVS does not include any color information.

\subsection{Periodogram's features}

If we want to recover the known classes of variable stars in an unsupervised context or improve accuracy in semi-supervised classification, it is necessary to design additional features that capture the main difference between light curves. Indeed, the period is an essential feature that can separate periodic variable star classes in supervised learning.  Numerous authors have suggested that a star's period is the most relevant feature to accurately classify variable stars using supervised algorithms \citep[e.g.,][]{dubath2011,richards2011,elorrieta2016,kim2016,jayasinghe2018}. However, in our unsupervised experiments,  the results were poor using the FATS features, the period, and the fitted Fourier parameters. These features does not provide sufficient information for our algorithms to group our data into variability classes, and in some cases, it introduces artificial clusters that essentially come from aliases peaks that are mistakenly chosen as best periods. Moreover, this error can propagate to the Fourier parameters calculated from the light curve, adding noise that results in even more abnormal clusters. Nevertheless, in large-scale surveys, the period will not give a meaningful measure to separate transient objects, quasi-periodic variables, some rotational variables, or eclipsing binaries (ECL), which motivates our search to proceed beyond the period and Fourier parameters as features themselves.

With this in mind, we realized that the information in the periodogram could potentially be exploited as is, without making assumptions about its maximum power overall. The periodogram stores information about main periodicities, harmonics, aliases, number of observations, and cadence. Subsequently, we followed a data-driven approach to extract significant peaks in a periodogram, using them to properly create features that describe its prominent periodicities and harmonics, which are the key to distinguish between variable objects. We had to be careful in finding a representation that minimizes the inclusion of cadence information. The following is the procedure that we devised to calculate, preprocess, and extract meaningful features from Lomb-Scargle periodograms \citep{periodogram1,periodogram2,periodogram3}:
\begin{enumerate}
    \item \textbf{Lomb-Scargle Periodograms:} The Lomb-Scargle periodograms are calculated using the \textsc{astropy} \citep{astropy:2013,astropy:2018} implementation, generating an equally spaced frequency grid from $\sim 0.0003\ \invdays$ to $24\ \invdays$, having $\sim 8\times 10^{5}$ evaluations.
    \item  \textbf{Peak detection:} The peaks were found through a peak detection algorithm from the Scipy library \citep{scipy}, \textit{find\_peaks}, dividing the periodogram in ten bins in the logarithm of the frequency. The parameter \textit{distance} (minimal horizontal distance, in terms of number of datapoints, between neighboring peaks) is set to the square root of each region's number of points. As a result, this effectively acts as a denoising procedure, rejecting  weaker peaks towards higher frequency bins. The average number of peaks is reduced from around $5\times 10^4$ to less than $10^3$, which can be easily stored for further experiments.    
    \item  \textbf{Aliasing filtering:} The typical aliases regions are carefully ignored. In the CSSCVS and OCVS cases, the synodic month alias is first removed, clipping about $0.001\ \invdays$ around it for all the stars. Second, we check if there is a sidereal day alias asserting that the first four one-day alias peaks are ordered decreasingly in power in a window of $0.05\  \invdays$ around each. Once they are detected, each such peak is removed up to its 24th harmonic. 
    For GDR2CVS, there is a known complex alias structure mainly explained by Gaia's $6\ \textrm{h}$ rotation period \citep{eyer2019}. It was removed, clipping $0.02\ \invdays$ around the frequencies that are multiples of $4\ \invdays$ up to $24\ \invdays$.
    \item \textbf{Log-transform:} The distribution of the periodogram's power is highly skewed. This has a detrimental effect when used in combination with other light curve features that are normally distributed. We applied a logarithm to the power to solve this issue, resulting in an approximately normal distribution and a narrower range of values.
    \item \textbf{Binning:} A binning is performed in frequency. A first edge is set at $0\ \invdays$. Then, a log-spaced sequence of seven edges is defined from $10^{-3}$ to $1\ \invdays$. Finally, a linear sequence of eleven edges is added from $1.5$ and $24\ \invdays$. This sums up to 18 bins. We take the greatest five maxima of the log-power from each bin ($\text{Max}_i$). This results in 18 features per maximum, totaling 90 features.
    \item \textbf{Scaling:} Finally, these features are scaled independently by their medians (Med) and IQRs for each of the five maxima (Max):
    \begin{equation}
    {\text{Max}'_{i}} = \frac{\text{Max}_i - \text{Med}_{\text{Max}_i}}{\text{IQR}_{\text{Max}_i}}, \ i=1,\cdots,5. 
    \end{equation}
    This scaling adjusts all these maxima to be comparable to each other. Still, further scaling of each feature will be necessary later on.
\end{enumerate}
 
\begin{figure*}
    \includegraphics[width=2\columnwidth]{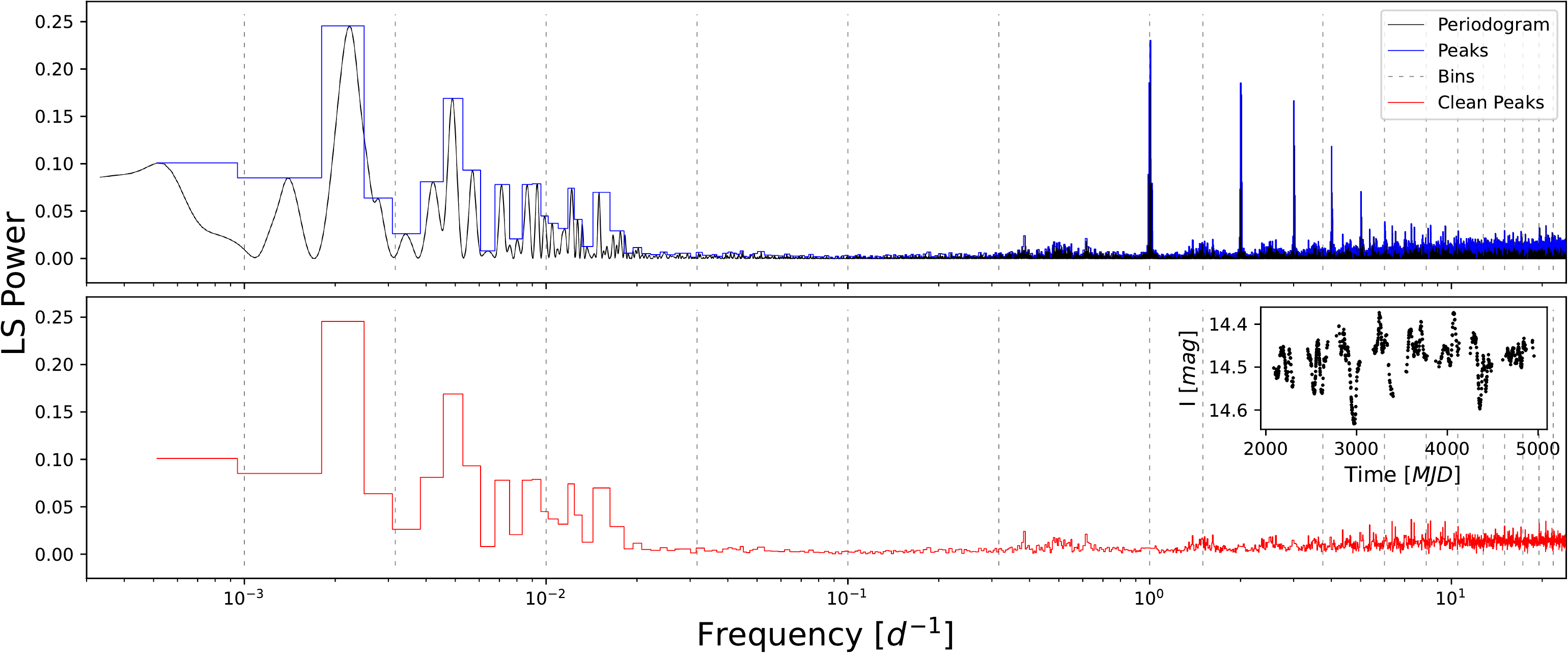}
    \caption{An example Lomb-Scargle periodogram of a semi-regular variable OGLE-SMC-LPV-11911 shown as a solid black line in the top panel. The peaks of this periodogram are drawn in blue steps. The alias-filtered peaks are shown in red steps at the bottom. Also, the bins from which we calculate the periodogram features are shown as vertical dashed lines. The inset plot at the bottom panel shows the original light curve as reference.}
    \label{periodogram}
\end{figure*}

An example using the semi-regular variable OGLE-SMC-LPV-11911 is shown in Figure~\ref{periodogram}. The periodogram is drawn in black in the top panel, while its $I$-band light curve is shown in the inset plot below. The detected peaks are drawn in blue steps at the top panel and the result of subsequent alias filtering in red steps at the bottom. The 18 bin edges are portrayed as gray vertical dashed lines. The figure shows that most of the peaks at lower frequencies are preserved while weak peaks are rejected gradually towards higher frequencies. Regarding the aliasing filtering, we see how the prominent one-day alias peaks were found and removed from this periodogram. We found a fair amount of stars having this one-day aliasing: $\sim 55\%$ in OCVS and $\sim 45\%$ in CSSCVS. Indeed, the classes most affected by one-day aliasing were long-period variables (LPV) and ECL. It is worthy of notice that a small number of stars with a one-day alias pattern did not match our formulation because they have only one one-day alias peak or other true signals overlap with certain peaks along the sequence of harmonics.

Wide bins are better to extract meaningful features since more than one characteristic peak associated with a specific class will have a high probability of overlapping in the same bin, thus leading to a decline of resolution. Hence, if the number of bins increases, these maxima will no longer be comparable in Euclidean distance for many classes of variable stars, because the maximum peaks (and their harmonics) will not be contained in the same bin. In experiments adopting hundreds of bins, we found that a tight binning creates a severely fragmented embedding, with intricated cluster shapes that are not easy to separate with clustering. Even with fewer bins, this representation can generate cluster fragmentation, especially for classes with a wide range in frequencies, e.g., LPVs. However, this fragmentation level was not too high to deter our clustering analysis.

The chosen scaling for the periodograms results in a more stable embedding than min-max normalization or others, blending well with the rest of the light curve features. Indeed, the min-max normalization was prone to group stars by observational cadence properties, since there is an explicit assumption about the maximum log-power in each bin and no assumptions about the periodogram's noise scale.

\subsection{Fourier features}

Using the calculated periodograms, conventional four-term Fourier components were determined by setting the minimum frequency given by the inverse of the light curve's baseline. We located the period at maximum power in our periodograms to fit a four-term Fourier model to the light curve. Then, the fitted residuals are used to compute new Lomb-Scargle periodograms so as to fit a new four-term Fourier model.

As described in \cite{debosscher2007}, we included as features the amplitudes, $A_{ij}$, and the phase components $\mathrm{PH}_{ij}$ for the two periods and each of the four correspondent harmonics ($\mathrm{PH}_{11}$ was not included since it was used as a reference phase). Also, the residuals of each fit can have relevant information about the presence of multi-periodicities. Thus, similarly as in \cite{dubath2011}, we use as features the ratio of the scatter of the residuals to the magnitudes, but using the IQR instead of the median absolute deviation for the two periodicities. This is done because the interquartile ranges are more robust when dealing with skewed distributions. Finally, we calculated the Abbe value of these residuals to measure their smoothness, in order to detect the existence of residual signals.

\begin{table*}
\caption{List of selected features used in this work}
\label{tab:features}
\begin{tabular}{>{\raggedright\arraybackslash}p{4.0cm}p{7cm}>{\centering\arraybackslash}m{3.0cm}}
    \hline
    \textbf{Feature} & \textbf{Description} & \textbf{Reference}\\
    \hline
    Robust Mean & Robust Mean measure of the magnitudes based on Huber's M-estimation & \cite{perez2017} \\ 
    MAD\starref & Median absolute deviation of the magnitudes & \cite{richards2011}  \\ 
	$Q_{31}$\starref & Difference between the $75^{th}$ and $25^{th}$ percentiles of the magnitudes &  \cite{kim2014} \\ 
	Robust Mean Variance\starref & Ratio of $Q_{31}$ to the Robust Mean magnitude & \dotso  \\ 
	Amplitude\starref & The median of the magnitudes over the 95th percentile minus the median of the magnitudes under the 5th percentile & \cite{richards2011} \\ 
	$R_{\rm CS}$\starref & Range of a cumulative sum of the magnitudes & \cite{kim2011} \\ 
	Beyond 1$\sigma$\starref & Percentage of points beyond 1$\sigma$ from the weighted mean & \cite{richards2011} \\ 
    Median BRP\starref & Fraction of points within a tenth of the magnitude range of the median magnitude & \cite{richards2011}  \\   
    Percent Amplitude\starref & Largest percentage difference between either the maximum or minimum magnitude and the median & \cite{richards2011} \\
    Upper outlier Fraction & The fraction of points over the 3rd quartile plus 1.5 times the interquartile range of the magnitude & This work \\ 
    GSkew\starref & Median based measure of the skew & \dotso  \\ 	    
    Flux Percentile Ratio Mid-20\starref & Sorted flux percentile ratio $F_{40, 60}/F_{5, 95}$& \cite{richards2011}  \\ 
    Flux Percentile Ratio Mid-35\starref & Sorted flux percentile ratio $F_{32.5, 67.5}/F_{5, 95}$& \cite{richards2011}  \\ 
    Flux Percentile Ratio Mid-50\starref & Sorted flux percentile ratio $F_{25, 75}/F_{5, 95}$& \cite{richards2011}  \\ 
    Flux Percentile Ratio Mid-65\starref & Sorted flux percentile ratio $F_{17.5, 82.5}/F_{5, 95}$& \cite{richards2011}  \\ 
    Flux Percentile Ratio Mid 80\starref & Sorted flux percentile ratio $F_{10, 90}/F_{5, 95}$& \cite{richards2011}  \\ 
    Percent Difference Flux Percentile\starref & Ratio of $F_{5, 95}$ over the median magnitude & \cite{richards2011}  \\ 
    Abbe Value & Measure of the smoothness of the light curve & \cite{vonneumann1941,vonneumann1942} \\	   
    Stetson$_K$\starref & Robust kurtosis measure based on Stetson variability index  & \makecell{\cite{stetson1996} \\ \cite{kim2011}} \\   
    Octile skewness (OS) & Robust measure of skewness & \makecell{\cite{brys2004} \\ \cite{perez2017}}\\ 	    
    Left octile weight (LOW) & Robust measure of the left tail weight & \makecell{\cite{brys2006}  \\ \cite{perez2017}}\\ 
    Right octile weight (ROW) &  Robust measure of the right tail weight & \makecell{\cite{brys2006}  \\ \cite{perez2017}}\\
    Robust Kurtosis &  Robust measure of kurtosis based on  on exceedance expectations & \cite{kim2004}\\	    
    Color & $V-I$ for OCVS; $G_{BP}-G_{RP}$ for GDR2CVS & \dotso  \\                           	
    Excess Abbe Value $T_{\rm sub}=50$~d & Estimation of the regularity of the light curve variability pattern for window size 50 days & \cite{mowlavi2014} \\	 
    Excess Abbe Value $T_{\rm sub}=100$~d & Excess Abbe Value for window size 100 days  & \cite{mowlavi2014} \\	
    Excess Abbe Value $T_{\rm sub}=250$~d & Excess Abbe Value for window size 250 days  & \cite{mowlavi2014} \\
    Slotted autocorrelation function length\starref & Robust autocorrelation function length for irregular time series & \cite{huijse2012}  \\
	$\textrm{Stetson}_K$ AC\starref & Stetson$_K$ applied over the slotted autocorrelation function  & \makecell{\cite{stetson1996} \\ \cite{kim2011}} \\
    QSO fit\daggerref & Quality of fit $\chi^{2}_{\rm QSO}/\nu$ for a quasar-like source, assuming $mag=19$ & \cite{butler2011}  \\
    QSO Null\daggerref & Natural logarithm of expected $\chi^{2}_{\rm QSO}/\nu$ for non-QSO variable. & \cite{butler2011} \\
    $\log(P)$ & Base 10 logarithm of the period & \dotso   \\
    $\Psi_{\rm CS}$\starref & $R_{\rm CS}$ applied to the phase-folded light curve & \cite{kim2014} \\ 
    $\Psi_{\eta}$\starref & Variability index $\eta_e$ applied to the the folded light curve & \cite{kim2014}\\ 
    $A_{ij}$ (8) & Amplitudes of the $j^{th}$ harmonic of the $i^{th}$ period & \cite{debosscher2007}\\
    $\log(R_{i1})$ (3) & Logarithm in base 10 of the amplitude ratios of the $j^{th}$ harmonic with respecto to the $1^{st}$ amplitude & \cite{debosscher2007} \\  
    $\mathrm{PH}_{ij}$ (7) & Phases of the $j^{th}$ harmonic of the $i^{th}$ period remmaped to be between $-\pi$ and $+\pi$ & \cite{debosscher2007}\\ 
    $\log($residuals-raw ratio$)$ (2)& Logarithm in base 10 of ratio between the IQR of the residuals of the fit periodic model and the IQR of the raw magnitudes & \cite{dubath2011} \\
    Residual's Abbe (2) & Abbe value of the residuals from the Fourier model subtraction of the first and second period & This work \\
	Periodogram $n^{th}$ maximums (90) & The first 5 maximums of the log-power for each the 18 bins & This work \\
  
    \hline
    \multicolumn{3}{l}{\footnotesize \textbf{Notes}: \starref~Feature from the FATS library; \daggerref~Feature from the Cesium library.}
\end{tabular}
\end{table*}

\section{Data visualization with \textsc{umap}}\label{sec:viz}

DR is a transformation done to the data to find a lower-dimensional embedding that approximately preserves its original structure or properties. DR is often used for reducing high-dimensional data for classification, visualization, feature selection, and/or feature extraction. Linear DR techniques such as Principal Components Analysis (PCA,~\citealt{pearson1901}), non-negative matrix factorization (NMF,~\citealt{patero1994}) or linear discriminant analysis (LDA,~\citealt{fisher1936}) have been extensively used. However, we know that datasets are often non-linear and more complex, so these linear DR algorithms are rarely the best choice. Fortunately, there are plenty of non-linear DR algorithms available (e.g., Isomap, \citealt{isomap}; SOM, \citealt{som}; t-SNE, \citealt{tsne}; Ivis, \citealt{ivis}), many of which are fairly proficient in handling our complex datasets.

Data visualizations can be very useful for understanding the data in general terms, enabling better strategies for further classification or clustering. In this section, we apply the \textsc{umap} algorithm for DR to two dimensions of our data for visualization (for a general description of the \textsc{umap} algorithm and its main parameters, refer to Appendix~\ref{appendix:umap}). First, we require scaling each feature to place all features approximately in the same range of values. In unsupervised learning, the effects of the scaling chosen are vast depending on the structure of the data. Most of the scaling strategies tested on our data resulted in an overlapping cluster structure or extreme fragmentation of clusters. Unsurprisingly, the best result was produced by one of the most ubiquitous scalings: subtracting the mean and dividing by the standard deviation of each feature independently. This scaling is known as standard scaling or Z-score, which centers each feature around zero and matches their standard deviation to a value of one.

\begin{figure*}
    \includegraphics[width=2\columnwidth]{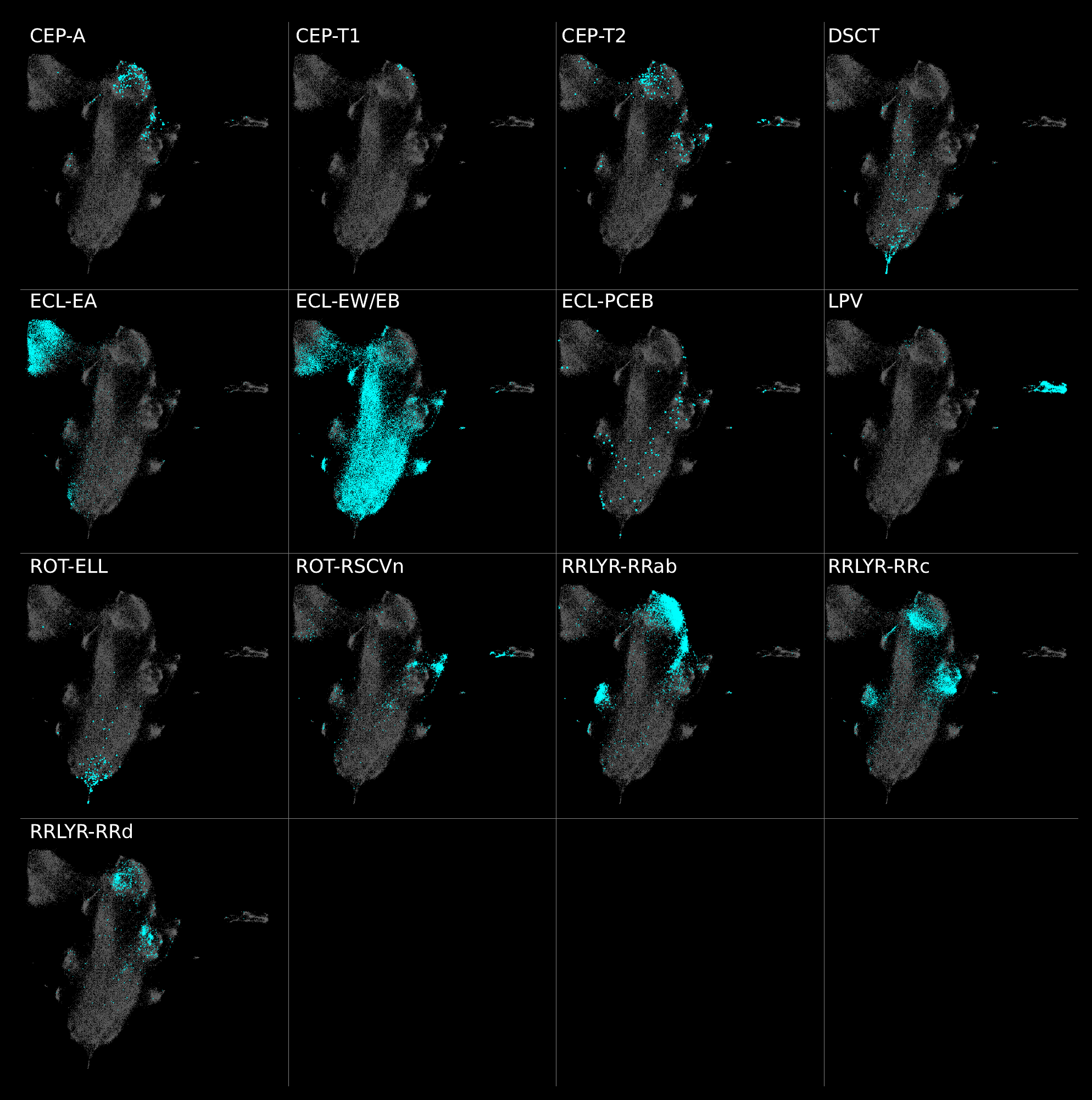}
    \caption{\textsc{umap} visualization of the CSSCVS. The embedding is drawn as gray points, with the different panels highlighting a specific sub-class (see Table~\ref{tab:catalogs}) in cyan.}
    \label{umap:csscvs}
\end{figure*}

\begin{figure*}
    \includegraphics[width=2\columnwidth]{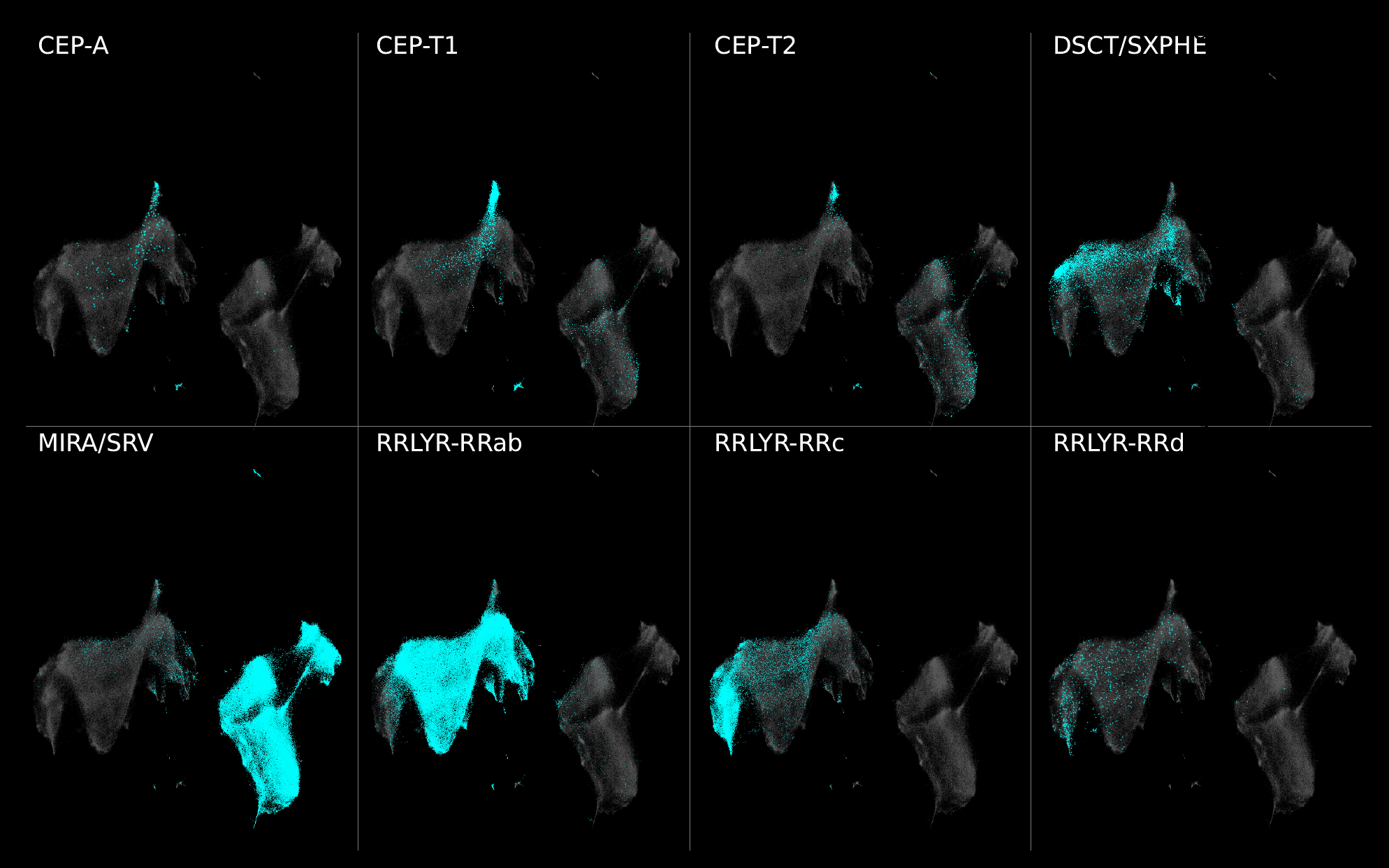}
    \caption{As in Figure~\ref{umap:csscvs}, but for the GDR2CVS case.}  
    \label{umap:gdr2cvs}
\end{figure*}

\begin{figure*}
    \includegraphics[width=2\columnwidth]{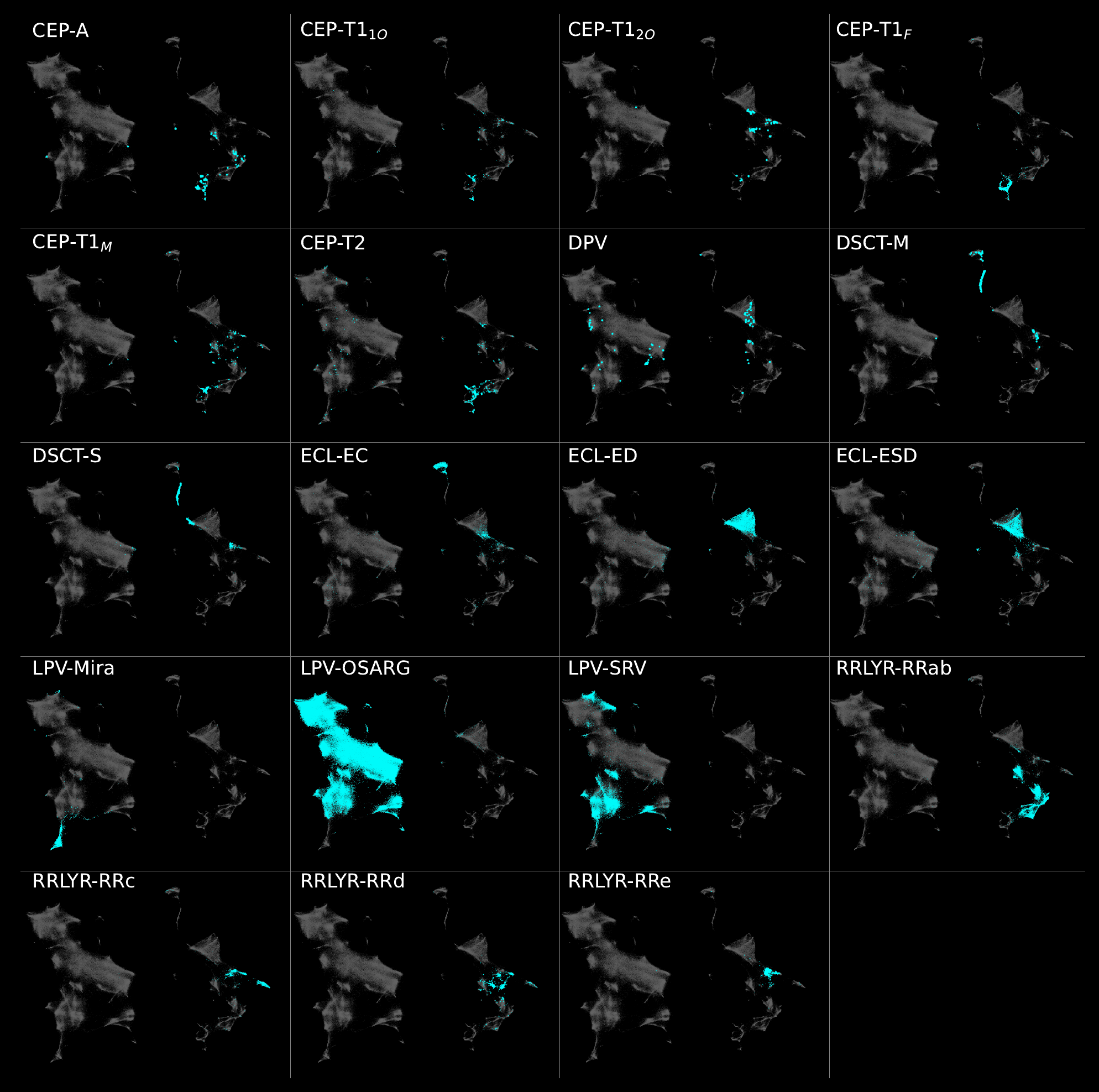}
    \caption{As in Figure~\ref{umap:csscvs}, but for the OCVS case.}
    \label{umap:ocvs}
\end{figure*}

After applying this scaling, we now can employ \textsc{umap} for DR to our catalogs. We reduced the data to two dimensions, so the  $\textit{n\_components}$ parameter was set to 2, the $\textit{n\_neighbours}$ parameter was set to 15, focusing on the local structure, and the $\textit{min\_dist}$ parameter to 0, creating compact structures. Figures~\ref{umap:csscvs},~\ref{umap:gdr2cvs} and~\ref{umap:ocvs} illustrate the derived embeddings for CSSCVS, GDR2CVS and OCVS respectively. In each panel, a sub-class is plotted in cyan over the embedding in gray. Also, the dot sizes were drawn larger for minority classes to improve their visibility. We can observe an evident separation between long-period variables (LPV) from other classes in these three embeddings, although some contamination endures. For the remaining classes, the difference between these visualizations becomes straightforward.

In the CSSCVS visualization (Fig.~\ref{umap:csscvs}), classes of variable stars appear to be in pure clusters but with a fair amount of cross-contamination, and the hierarchical structure of clusters does not seem to have an obvious explanation. There may be one main reason for this: the smaller size of these data. The data's local structure will be accurately represented by the weighted k-nearest neighbors graph built by \textsc{umap} as long as a sufficient number of samples is present in that region (defined by a sub-class). Having enough samples of a certain class becomes more critical for classes that are intrinsically difficult to distinguish between each other. Ultimately, this graph is the key to represent the data accurately, so in this context, fewer examples per sub-class might imply uncertain or noisy embedding. In contrast, GCVS visualization displays detached clusters of sub-classes at the non-LPV region. However, there is high cross-contamination between some of these clusters. This visualization represents the opposite case of CSSCVS: well-populated sub-classes and fewer observations per light curve.

Finally, OCVS visualization (Fig.~\ref{umap:ocvs}) has the largest apparent separations between classes, which we believe is explained by combining the light curves' high number of observations and the fair amount of samples per sub-class. We can observe outliers, clusters of classes, or sub-classes and over-dense regions inside these clusters.

In these visualizations, some clusters match sub-classes, and some fragmentation is observed in many clusters (e.g., RRLYR-RRab, CEP, LPV-SRV). Certainly, there are many factors in play to form these sub-structures, including periodogram similarity, the presence of unfiltered one-day aliases, or the intrinsic noise of some features.

\section{Clustering Analysis}\label{sec:clustering}

Clustering algorithms allow us to study the underlying structure of the data, finding groups or partitions that have properties in common. Clustering analysis itself is an obscure task because it is difficult to define what a cluster is. This, in part, explains why there are many clustering algorithms based on diverse approaches, such as partitioning the feature space \citep[e.g., K-Means,][]{kmeans} or through measuring local density \citep[e.g., \textsc{dbscan},][]{ester1996}. Depending on the algorithm, some prior knowledge about the data could be required, e.g., the number of clusters or the minimum size of a cluster in samples (i.e., stars) . Also, the criteria used to evaluate how accurate the partitions are can be difficult to define and are delimited by the task's goal.

In large-scale surveys, these complications are mitigated somewhat since we broadly know what objects we expect to find. Moreover, we could perform cross-matching to evaluate which classes or sub-classes of variable stars are present in the data. This prior knowledge of the data can be used effectively to propose a clustering analysis method applicable to any survey. 
For this procedure, we adopted \textsc{umap} for DR and visualization and the \textsc{hdbscan} \citep{hdbscan} algorithm for clustering (refer to Appendix~\ref{appendix:hdbscan} for a general description of \textsc{hdbscan} and its main parameters). This analysis aims to explore whether it is possible to recover pure clusters of classes or sub-classes and report other rare aggregations. In the following subsections, we describe our clustering analysis procedure and then show the results of implementing it with our catalogs.

\subsection{Clustering analysis procedure}\label{sec:procedure}

Clustering analysis is, for most applications, an exploratory task. Thus, we have to define clear goals and methods to set exactly what we pursue in this endeavor. As the main goal, we want to find high-density clusters in the data that are completely or partially isolated. We expect that these clusters correlate with the classes or sub-classes of variable stars.

We propose a hierarchical clustering procedure: a combination of \textsc{umap} DR and the \textsc{hdbscan} algorithm to embed the data and cluster it into smaller structures at each level. \textsc{umap} helps us visualize the sub-sets of the data, gaining valuable insights about its structure and transforming the data to a simplified lower-dimensional version of it while maintaining most of its properties, which improves the results of \textsc{hdbscan}. Furthermore, a hierarchical clustering procedure makes our analysis simpler, dividing the task into smaller ones. Our datasets usually form clusters of various shapes and properties, so it is difficult to find \textsc{hdbscan} parameters that fully capture these structures. Finding \textsc{hdbscan} parameters becomes more manageable with our method since we focus on smaller structures that will become progressively more homogeneous at each iteration, i.e., clusters of similar densities. The general procedure can be outlined as follows:
\begin{enumerate}
    \item Use \textsc{umap} to reduce the data to 20 dimensions for \textsc{hdbscan} clustering and to 2 dimensions for visualization.
    \item Use \textsc{hdbscan} over the 20-dimensional data to find large structures (an aparent set of clusters grouped together) and small clusters at the top of the hierarchy. 
    \item Perform \textsc{umap} again for each of the large structures that could potentially be partitioned further.
    \item Perform \textsc{hdbscan} on each large structure forming new hierarchies when necessary.
\end{enumerate}

We realized no more than two or three hierarchy levels to cluster the data. This choice depends on the amount of data and the number of classes or sub-classes in these data.  It should be noted that the \textsc{hdbscan} clustering should be done over the 20-dimensional data and not over the two-dimensional visualization since some clusters are not well represented in the latter. 

Our clustering analysis procedure is quite straightforward, nevertheless we followed a strategy to be concise and consistent in our analyses:
\begin{itemize}
    \item[--] We intend to capture isolated groups or dense regions inside large structures in the data. Thus, visualizations can aid us in choosing appropriate \textsc{hdbscan} parameters. Moreover, the visualization of subsets of the data usually gives us more information on its fine structure. If this subset is one of the clusters resulting from \textsc{hdbscan} clustering, then some noise was removed at that stage, which improves the result of \textsc{umap} visualization.

    \item[--] We should capture as many of the small clusters displayed in the visualization as possible. These could be astrophysically interesting objects (e.g., transients, binary systems in which one or more of the components is itself a variable) or artifacts (e.g., blends). The smallest clusters may have only a handful of members, so their light curves could be analyzed by visual inspection.

    \item[--] The first clustering generated over our 20-dimensional data should expose very large structures and other extreme outliers at the top of the \textsc{hdbscan} hierarchy. In this respect, we are searching for a small amount of clusters while we are trying to minimize the number of stars assigned as noise by \textsc{hdbscan} (small $min\_samples$, see Appendix~\ref{appendix:hdbscan}).

    \item[--] After the first hierarchy level, if \textsc{hdbscan} parameters do not provide good results, a new hierarchy should be created to find the smallest clusters plus one or two large structures. Then, \textsc{hdbscan} should be run again on the large structures that could be split further. In our analyses, this occurs when there are several clusters of various sizes and densities in the data, and the parameters to find them, without dropping too much data as noise, were impossible to recover.
\end{itemize}

\subsection{Clustering evaluation}

We calculated evaluation metrics for each cluster or sets of clusters to draw quantitative conclusions about the results. Subsequently, we use just one simple external measure, purity. Given a cluster $\omega$, the purity $\mathcal{P}(\omega)$ is defined as the fraction of the most frequent class contained in the cluster, or
\begin{equation}
     \mathcal{P}(\omega) = \frac{1}{n_\omega}\max_{c}^{C}{\{\omega_c\}}, 
\end{equation}
\noindent where $n_\omega$ is the number of points in the cluster, $\omega_c$ is the number of points of class c, and C is the set of all classifications. We can extend this metric for a set of $k$ clusters ${\Omega}= \{\omega_1,\omega_2,...,\omega_k\}$ as follows:
\begin{equation}
     \mathcal{P}(\Omega) = \frac{1}{N} \sum_{1}^{k} n_{\omega_i} \mathcal{P}(\omega_i), 
\end{equation}
\noindent where $N = \sum_{1}^{k} n_{w_i}$ is the total number of points in $\Omega$. In other words, the purity $\mathcal{P}(\Omega)$ is equivalent to an average of cluster purities $\mathcal{P}(\omega_i)$, weighted by the size of each cluster. 
In our analyses, we will measure the cluster purity $\mathcal{P}(\omega)$ respect to classes and sub-classes to gain insights on how the purity distributes among them. Also, we will compute $\mathcal{P}(\Omega)$ of a set of clusters of assigned sub-classes to comprehensively quantify the limitations and effectiveness of our clustering procedure.

\subsection{Results on CSSCVS}

\begin{figure*}
    \centering
    \includegraphics[width=2\columnwidth]{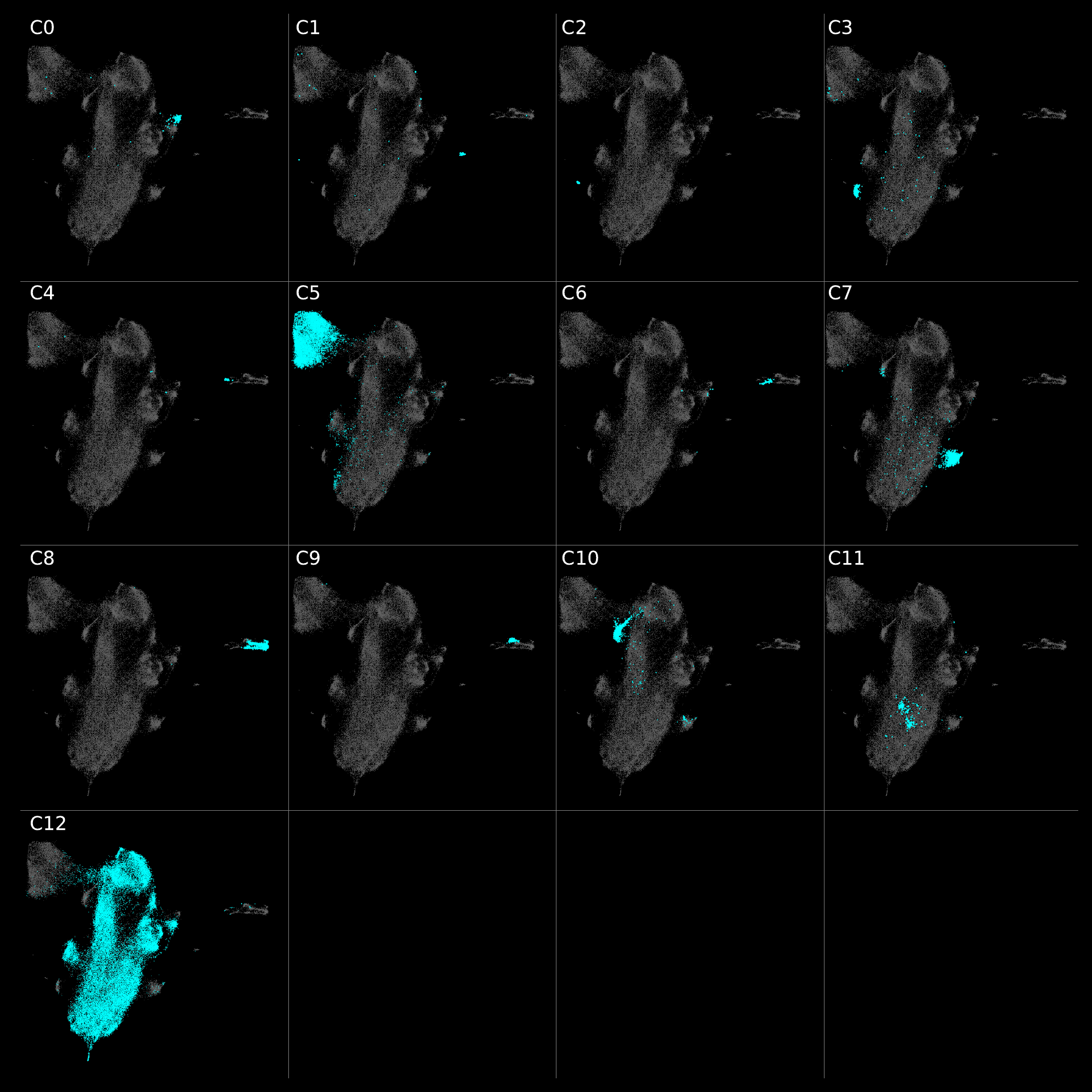}
    \caption{Results on the first \textsc{hdbscan} clustering done to CSSCVS. The embedding is colored in gray for reference, 
    with each panel highlighting a specific cluster in cyan.}
    \label{clust:csscvs1}
\end{figure*}

The CSSCVS visualization in Figure~\ref{umap:csscvs} depicts LPVs and some ROT-RSCVns and ECL-EAs being far away from the central structure. This indicates dissimilarity between these clusters, which was confirmed when we performed \textsc{hdbscan} clustering on these data. These clusters are stable to changes in the parameters, i.e., it is difficult to obtain a clustering with these clusters being partitioned differently. Figure~\ref{clust:csscvs1} shows the clustering results at this hierarchy level, with each cluster highlighted in cyan and the embedding depicted in gray for reference. By comparing with Figure~\ref{umap:csscvs}, we see that we found clusters of LPVs (C8 and C9), ROT-RSCVn (C0, C4, and C6), and EA (C5). Also, there are other groups of binaries that are separated from the central C12 cluster. Cluster C10 essentially contains ECL-EW/EB stars with periods around 0.35~d with a similar periodogram harmonic pattern, and a high-purity cluster C7 contain mainly ECL-EW stars.

\begin{figure*}
    \centering
    \includegraphics[width=2\columnwidth]{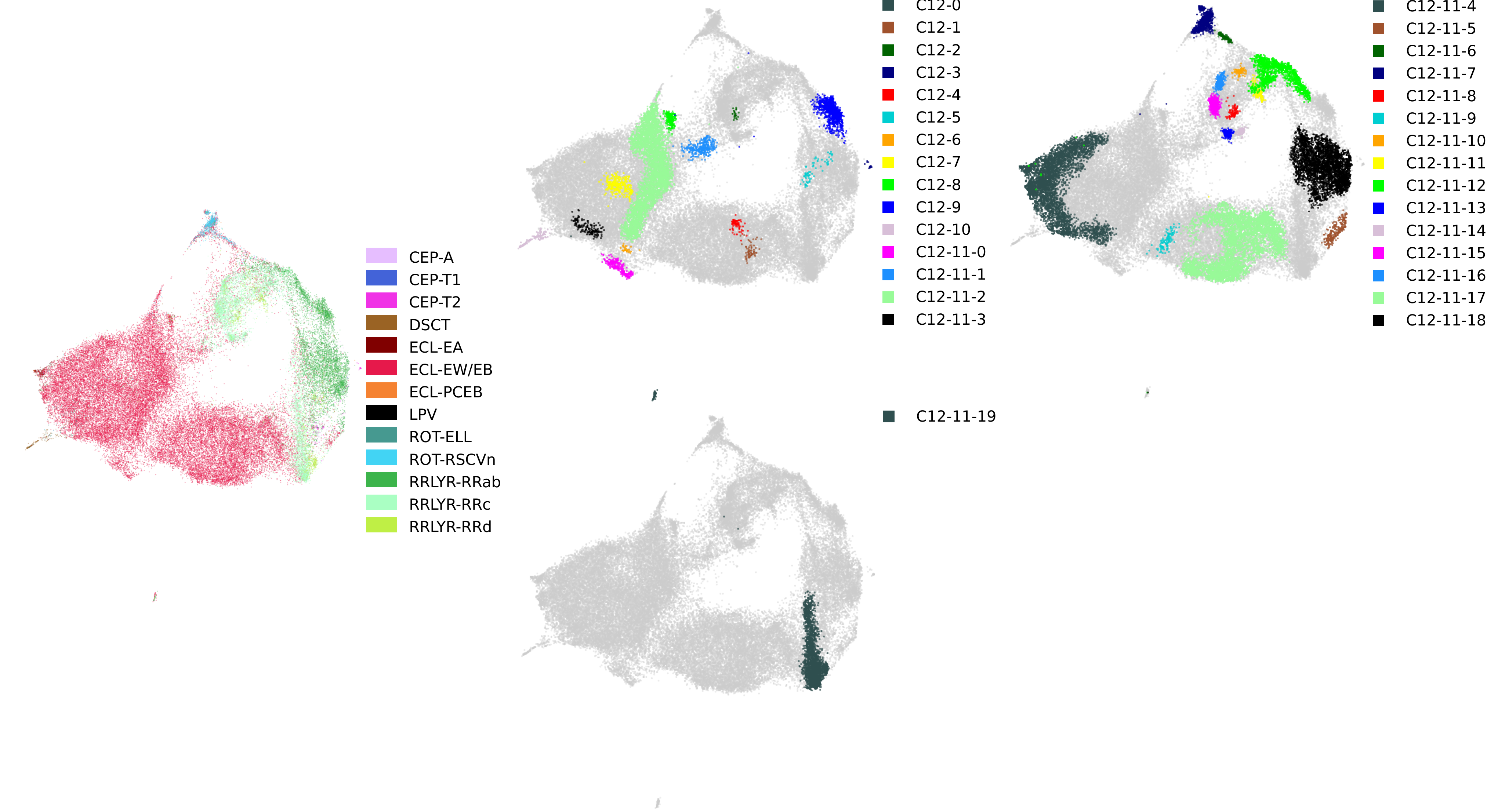}
    \caption{The left panel is a \textsc{umap} visualization of C12 from CSSCVS. The results on the \textsc{hdbscan} clustering done to cluster C12 from CSSCVS is shown in the three panel at the right. Gray points in the latter panels are drawn for reference purposes only, according to the full C12 data shown in the left panel.}
    \label{clust:csscvs2}
\end{figure*}

Following the procedure described in Section~\ref{sec:procedure}, we executed \textsc{umap} on the raw data of the stars in cluster C12, which presents a nested inner structure. Figure~\ref{clust:csscvs2} shows a \textsc{umap} visualization of cluster C12 in the left panel. The results of the C12 clustering is divided in three panels to improve the visualization. Also, gray reference points of the embedding were added for easy comparison. The nested nature of C12 requires clustering to be done at two additional levels, cutting the condensed tree at a small $\hat{\epsilon}$ (see Appendix~\ref{appendix:hdbscan}) using the leaf method. For this reason, at the bottom level, we find that, the large substructure C12-11 may be comprised of an additional set of 20 substructures.

\begin{figure}
    \centering
    \includegraphics[width=1\columnwidth]{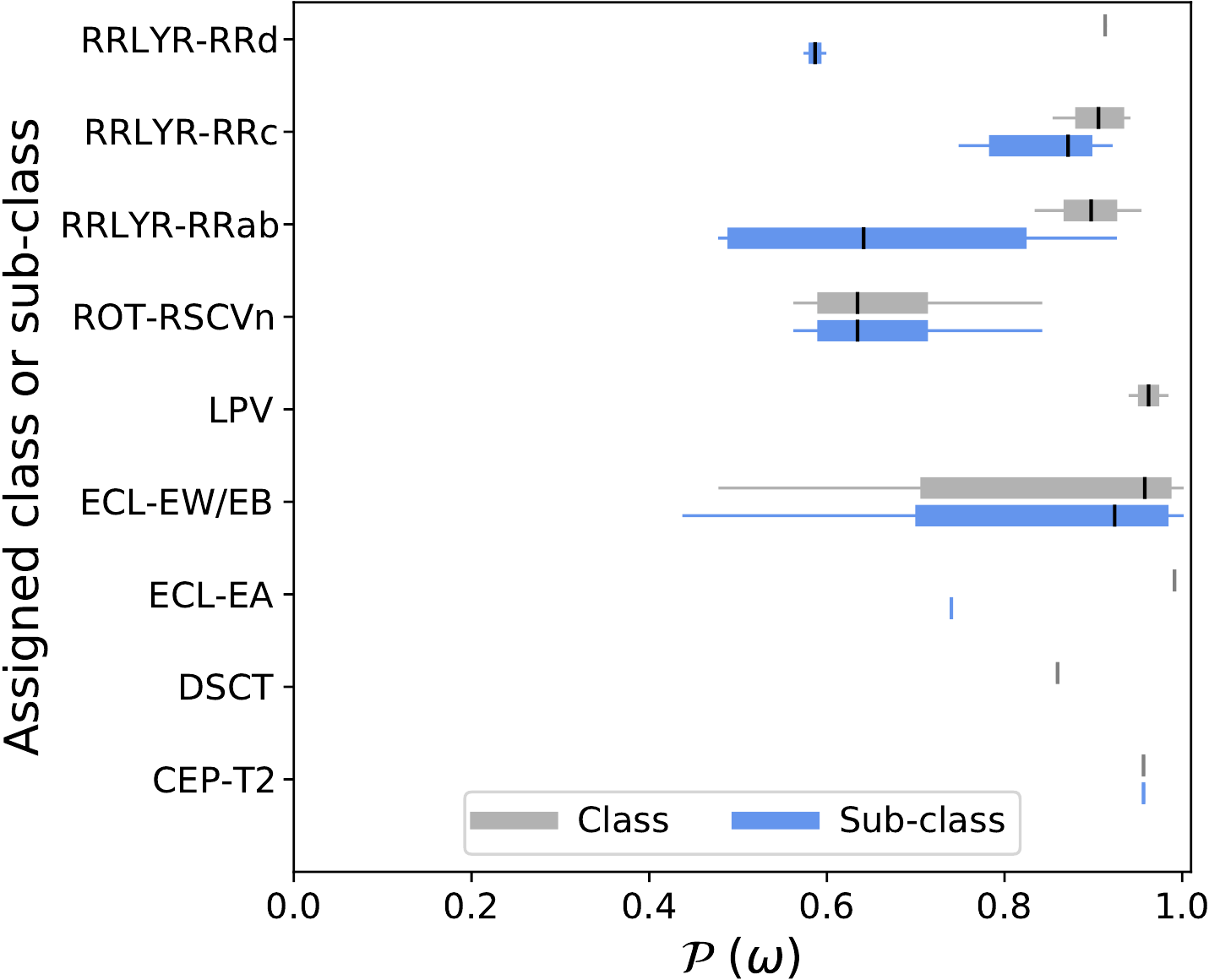}
    \caption{Cluster purity $\mathcal{P}(\omega)$ box-plots measured by class (gray) and sub-classes (light blue) assigned for CSSCVS. The black vertical lines indicate the median. If only one cluster is found, a vertical gray or light blue line is drawn instead.}
    \label{purities:csscvs}
\end{figure}

As a result, we partitioned about 54\% of the data (the rest is assigned as noise by \textsc{hdbscan}) into 43 clusters. Figure~\ref{purities:csscvs} shows box-plots for the results on the purity $\mathcal{P}(\omega)$ of these 43 clusters found for each assigned sub-class measured by classes (in gray) and sub-classes (in light blue). We show a black line to indicate the median, and we do not show the boxes when only one cluster was found, which is the case of RRLYR-RRd, ECL-EA, DSCT, and CEP-T2. We observe a clear difference for some sub-classes when the purity is measured for classes or sub-classes.

Table~\ref{tab:purities} shows the results on the purity $\mathcal{P}(\Omega)$ for clusters of the assigned sub-classes. Purity is not shown in case there are not sub-classes available. What stands out in this table is that purity measured by class reaches near 0.9 for most clusters, and it drops prominently when it is measured with respect to sub-classes. We have to interpret this with caution, considering that this catalog does not provide sub-classes for some classes, such as LPVs.

\subsection{Results on GDR2CVS}

\begin{figure*}
    \centering
    \includegraphics[width=2\columnwidth]{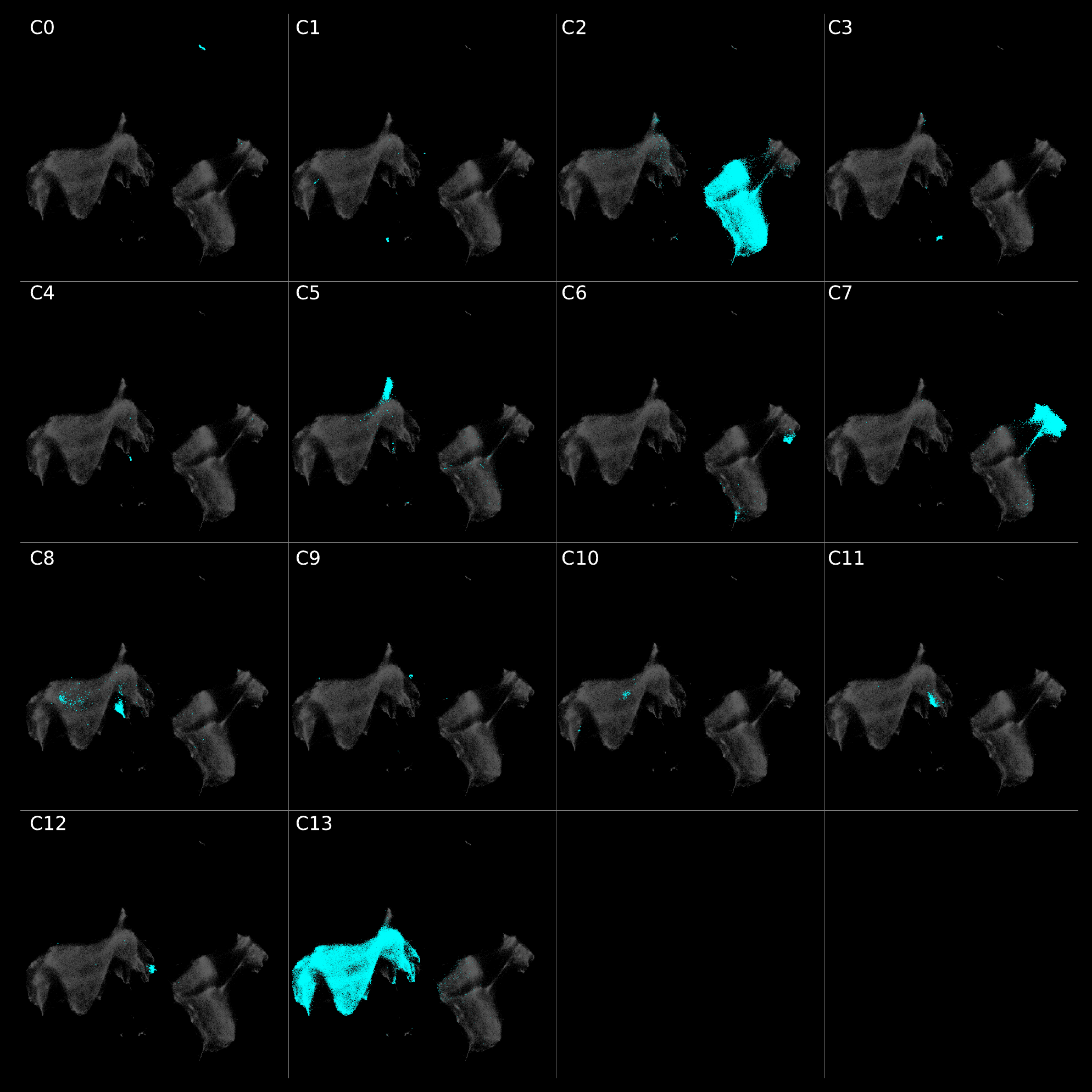}
    \caption{As in Figure~\ref{clust:csscvs1}, but for GDR2CVS.}
    \label{clust:gdr2cvs1}
\end{figure*}

For this catalog, whose initial visualization can be found in Figure~\ref{umap:gdr2cvs}, our clustering analysis reveals a complex structure made of noisy nested clusters and outliers. Although many of these clusters belong to a particular class, most of them have properties that separate them from the rest of the data, e.g., linear trends or total number of observations below 18. The first \textsc{hdbscan} clustering on this catalog is shown in Figure~\ref{clust:gdr2cvs1}. Many small isolated clusters were found, and clusters of over 0.9 in purity with respect to classes such as CEP (C3 and C5) and LPV (C2, C6, C7).

Cluster C0 lies isolated in the upper part of the visualization. It contains LPV stars that have around 12 to 18 observations, with apparently incomplete phase coverage. The same happens with clusters C1, C4 and C8 which have around 20 noisy observations each, making it very difficult to confirm their association to actual variability classes. On the other hand, clusters C9, C10, C11, and C12 have anomalous high-power peaks resulting from a failed Lomb-Scargle periodogram calculation, caused by a small number of observations (around 15 each). It is worth mentioning that some light curves in clusters C6 and C7 appear to correspond to Mira variables with their characteristic long periods and sinusoidal light curves.

\begin{figure*}
    \includegraphics[width=2\columnwidth]{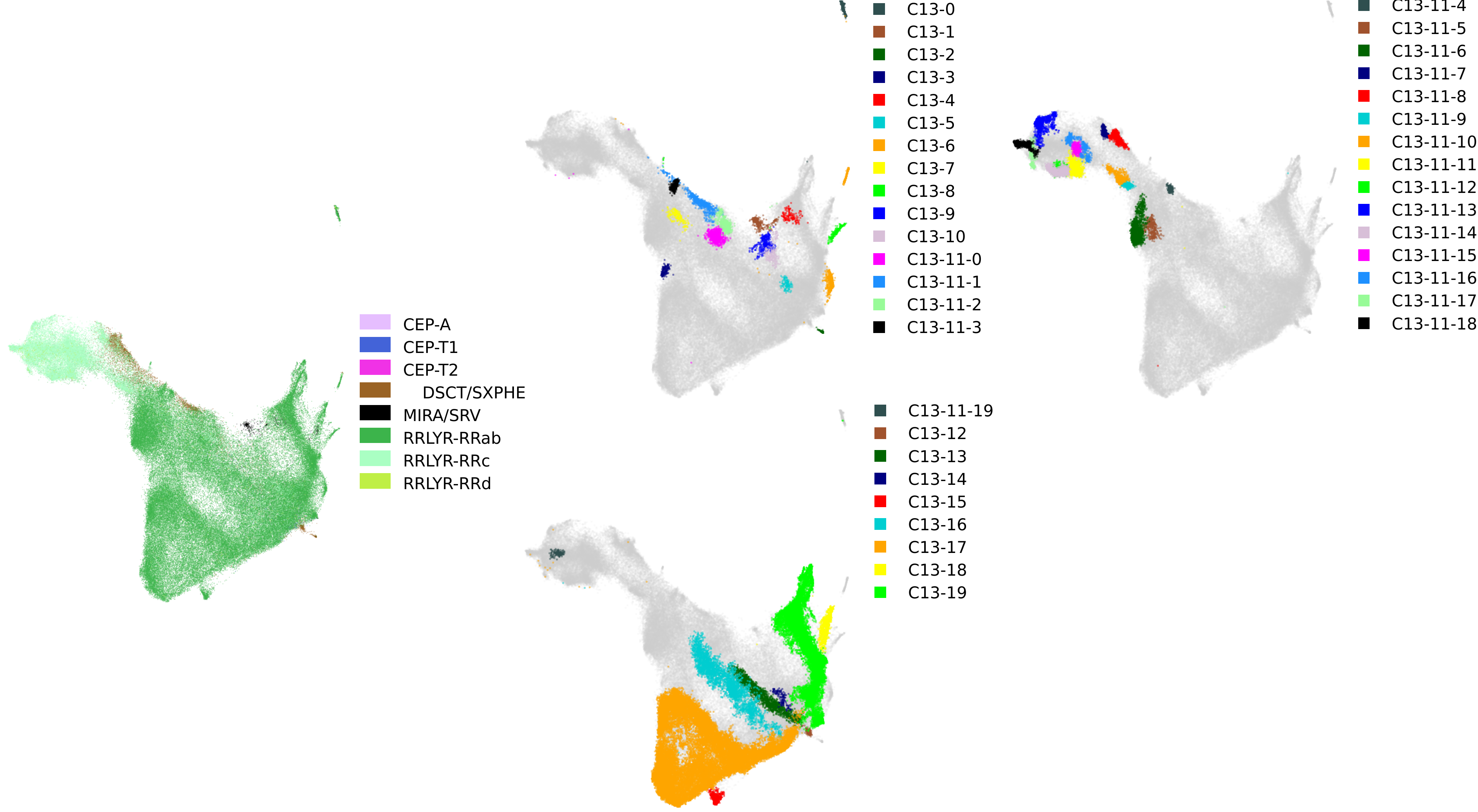}
    \caption{The left panel is a \textsc{umap} visualization of C13 from GDR2CVS. The results on the \textsc{hdbscan} clustering done to cluster C13 from GDR2CVS is shown the three panel at the right. Gray points in the latter panels are drawn for reference purposes only, according to the full C13 data shown in the left panel.}
    \label{clust:gdr2cvs2}
\end{figure*}

The next step is to take the large cluster C13 to perform a \textsc{umap} DR and \textsc{hdbscan} clustering analysis. The results are shown in Figure~\ref{clust:gdr2cvs2}. As done previously, the visualization with true labels is shown at the left, and the result of the clustering is shown on the three panels at its right. 

Cluster C13 is difficult to partition since classes spread and merge in different ways. We had to cluster C13 at two levels since it contains subclusters of various sizes and a large group of RRLYR-RRc and DSCT/SXPHE stars (C13-11). The visualization shows that RRab stars bind all the clusters together, contaminating the RRc and DSCT/SXPHE clusters. Indeed, there is a mix of pure and contaminated clusters in C13. RRLYR-RRc variables heavily pollute the largest DSCT/SXPHE cluster (C13-11-8); however, we found pure clusters (C13-2 and C13-12) of DSCT/SXPHE that lie just outside the  RRab cluster.

\begin{figure}
    \centering
    \includegraphics[width=1\columnwidth]{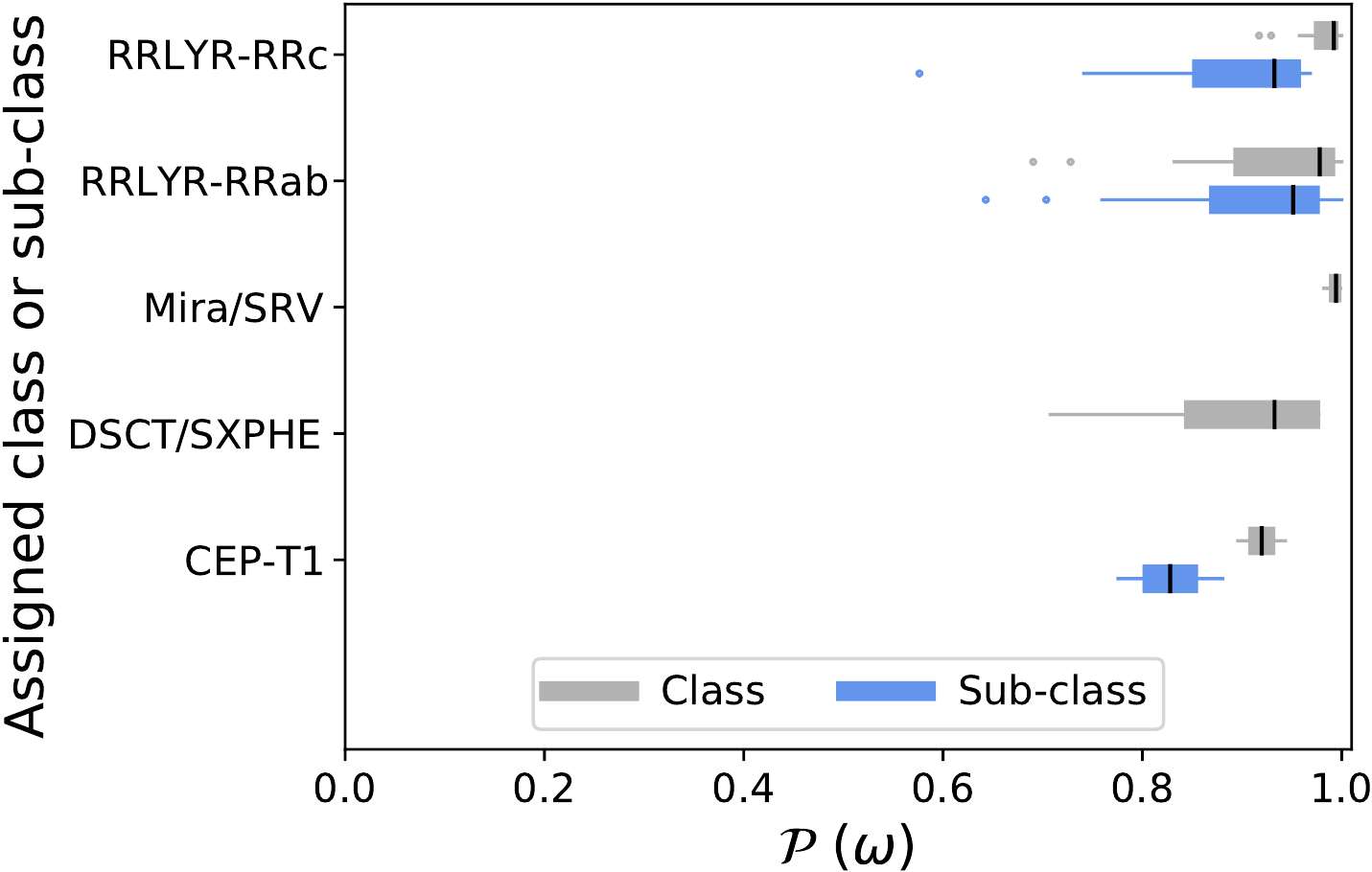}
    \caption{Cluster purity $\mathcal{P}(\omega)$ box-plots measured by class (gray) and sub-classes (light blue) assigned for GDR2CVS. The black vertical lines indicate the median.}
    \label{purities:gdr2cvs}
\end{figure}

Overall, we clustered near 63\% of the data into 52 clusters. Metrics for the result on this clustering are also shown in Table~\ref{tab:purities}. In Figure~\ref{purities:gdr2cvs}, we show the results on the purity of the clusters found. From the figure, it can be seen that most of the clusters have high purity. In Table~\ref{tab:purities}, we see that most of the sub-classes have high purity except for DSCT/SXPHE, which has a large cluster with purity near 0.70 (C13-11-1). Although we found pure CEP clusters (purity higher than 0.9), we could not partition this cluster further into sub-classes.

\subsection{Results on OCVS}

\begin{figure*}
    \centering
    \includegraphics[width=2\columnwidth]{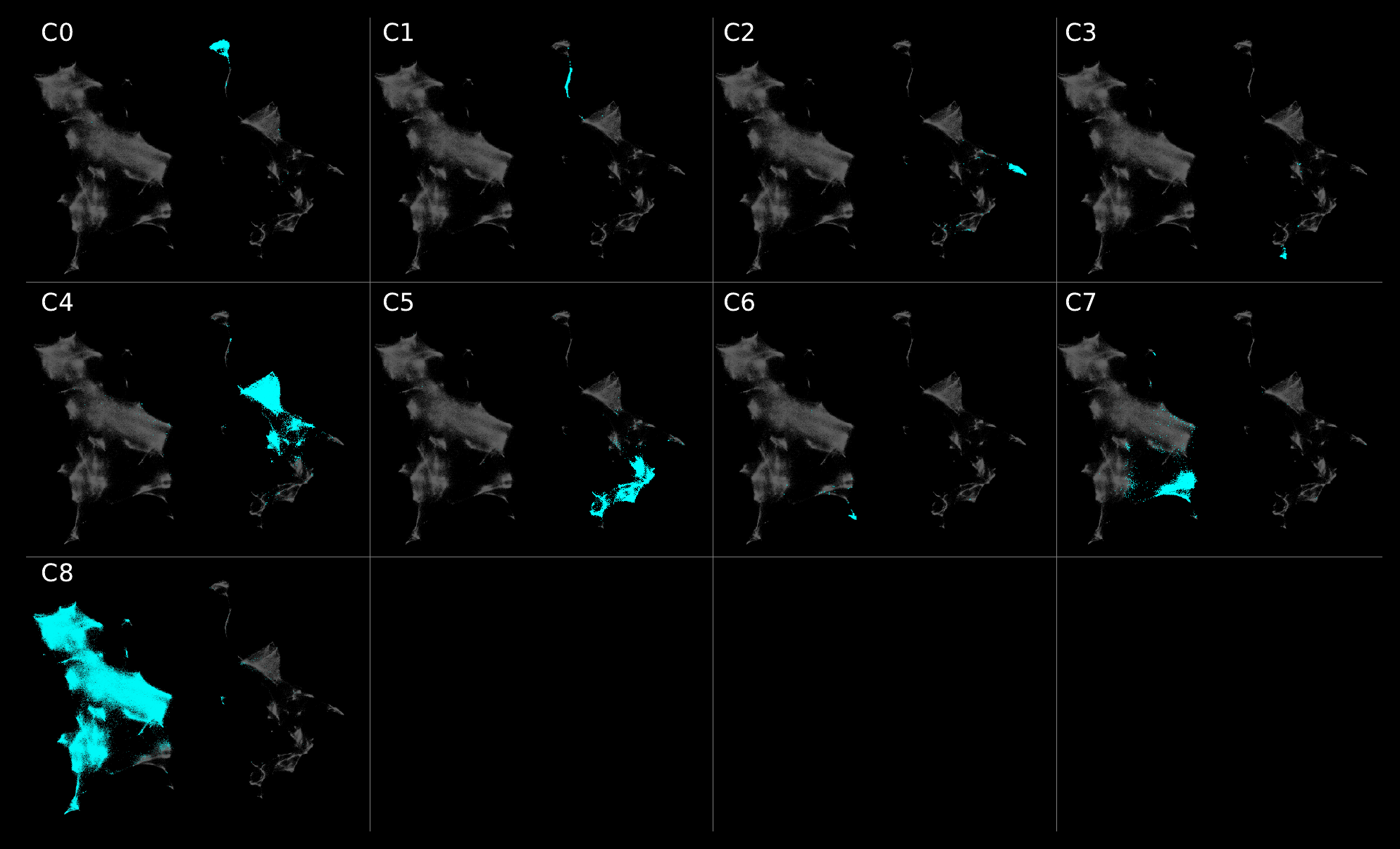}
    \caption{As in Figure~\ref{clust:csscvs1}, but for OCVS.}
    \label{clust:ocvs1}
\end{figure*}

The OCVS data contains many nested clusters inside large structures that are well defined. The first division was performed so as to group these large structures as shown in Figure~\ref{clust:ocvs1} (compare with Fig.~\ref{umap:ocvs}). In this process, some compact and highly pure clusters were found, such as C0 (ECL-EC with 0.97), C1 (DSCT with 0.95), C2 (RRLYR-RRc with 0.98), and C3 (CEP-T$_{\rm 1O}$ with 0.94). 

\begin{figure*}
    \includegraphics[width=2\columnwidth]{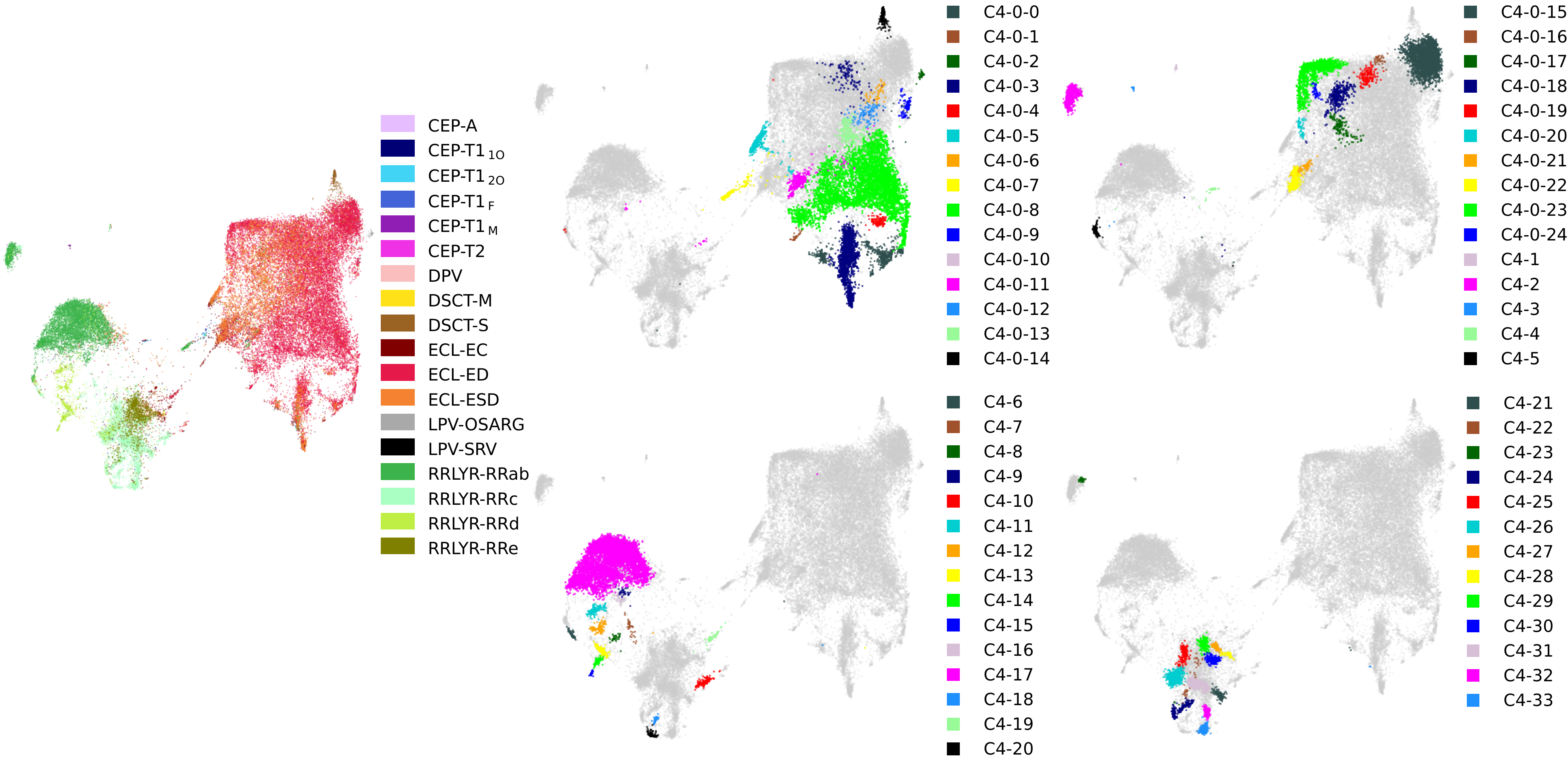}
    \caption{The left panel is a \textsc{umap} visualization of C4 from OCVS. The results on the \textsc{hdbscan} clustering done to cluster C4 from OCVS is shown in the four panel at the right. Gray points in the latter panels are drawn for reference purposes only, according to the full C4 data shown in the left panel.}
    \label{clust:ocvs2}
\end{figure*}

We now proceed with the analysis of clusters C4 (DSCT, ECL and RRLYR), C5 (CEP and RRLYR), and C8 (LPVs). Figure~\ref{clust:ocvs2} presents the visualization for cluster C4 at the left and the results of the clustering in the four panels at its right. This cluster has a clear separation between classes but noisy borders separating sub-classes. However, sub-classes like CEP-T1$_{\rm M}$ (C4-3) or RRLYR-RRd (C4-11 to C4-15) have a purity over 0.9. Some variables have clusters of a wide range of purities, such as RRLYR-RRe (C4-22 and C4-29 with around 0.75) and DSCT (C4-0-14 with 0.98, C4-27 with 0.72, and C4-28 with 0.54). This is explained by the mixture of many low amplitude variables of different classes and/or sub-classes in this cluster.

As it can be seen, this clustering was performed in two steps, subdividing cluster C4-0 (mostly ECL) into smaller clusters. Contrary to the other classes in C4, the ECL class does not present any explicit pure clusters of sub-classes, but only a smooth transition between ECL-ED to ECL-ESD, clearly depicted in the visualization.

\begin{figure*}
    \includegraphics[width=2\columnwidth]{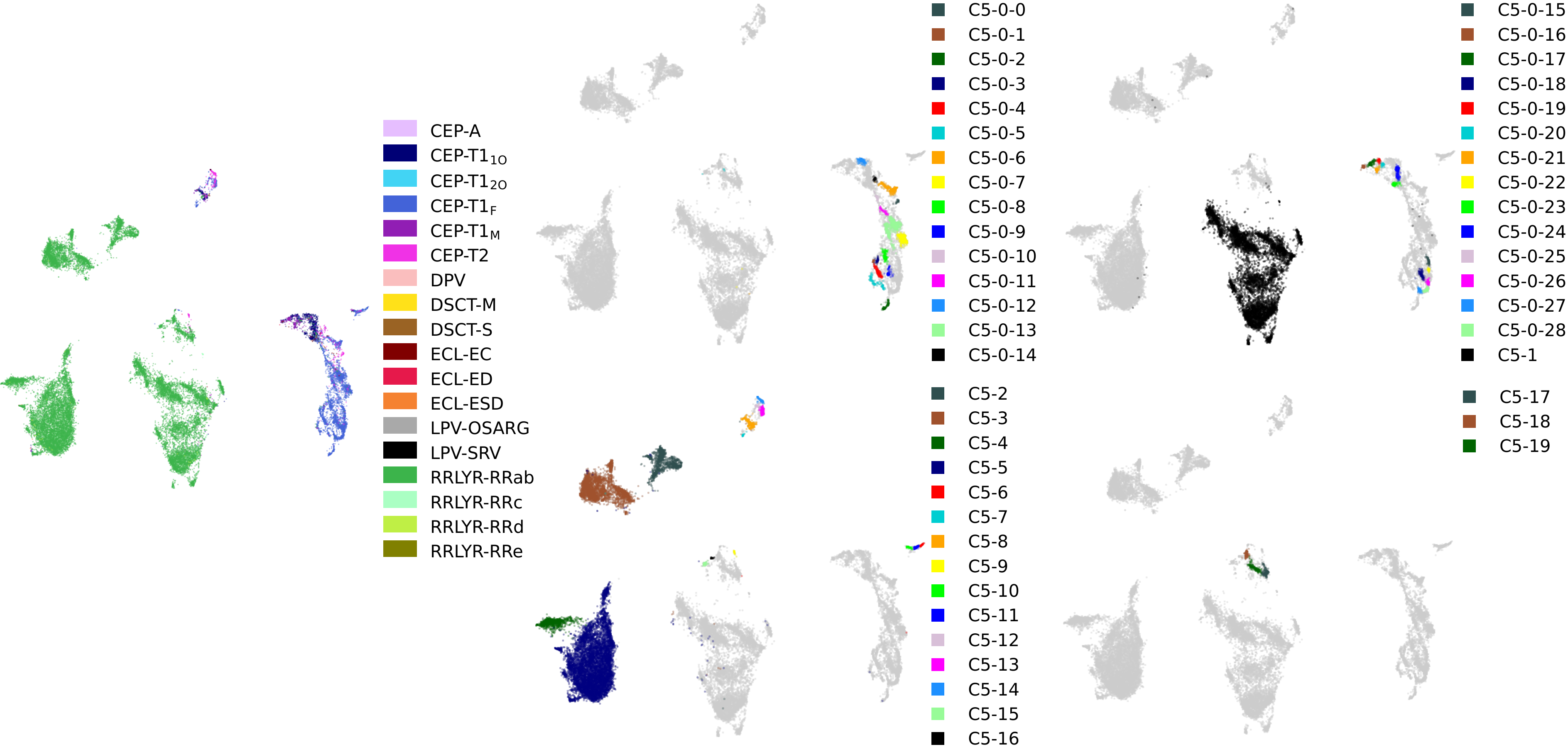}
    \caption{The left panel is a \textsc{umap} visualization of C5 from OCVS. The results on the \textsc{hdbscan} clustering done to cluster C5 from OCVS is shown in the four panel at the right. Gray points in the latter panels are drawn for reference purposes only, according to the full C5 data shown in the left panel.}
    \label{clust:ocvs3}
\end{figure*}

The visualization for cluster C5 is shown in the left panel in Figure~\ref{clust:ocvs3}, and the results on clustering are shown in the four panels at its right. Note that this visualization has a more apparent separation between the classes as compared to the original one. The clustering here was done in two steps to subdivide the CEP cluster (C5-0). In this case, RRLYR-RRab and CEP-T1$_{\rm F}$ clusters have an average in purity of 0.98 and 0.96 respectively. Other CEP sub-classes are mixed and do not show isolated clusters.

\begin{figure*}
    \includegraphics[width=2\columnwidth]{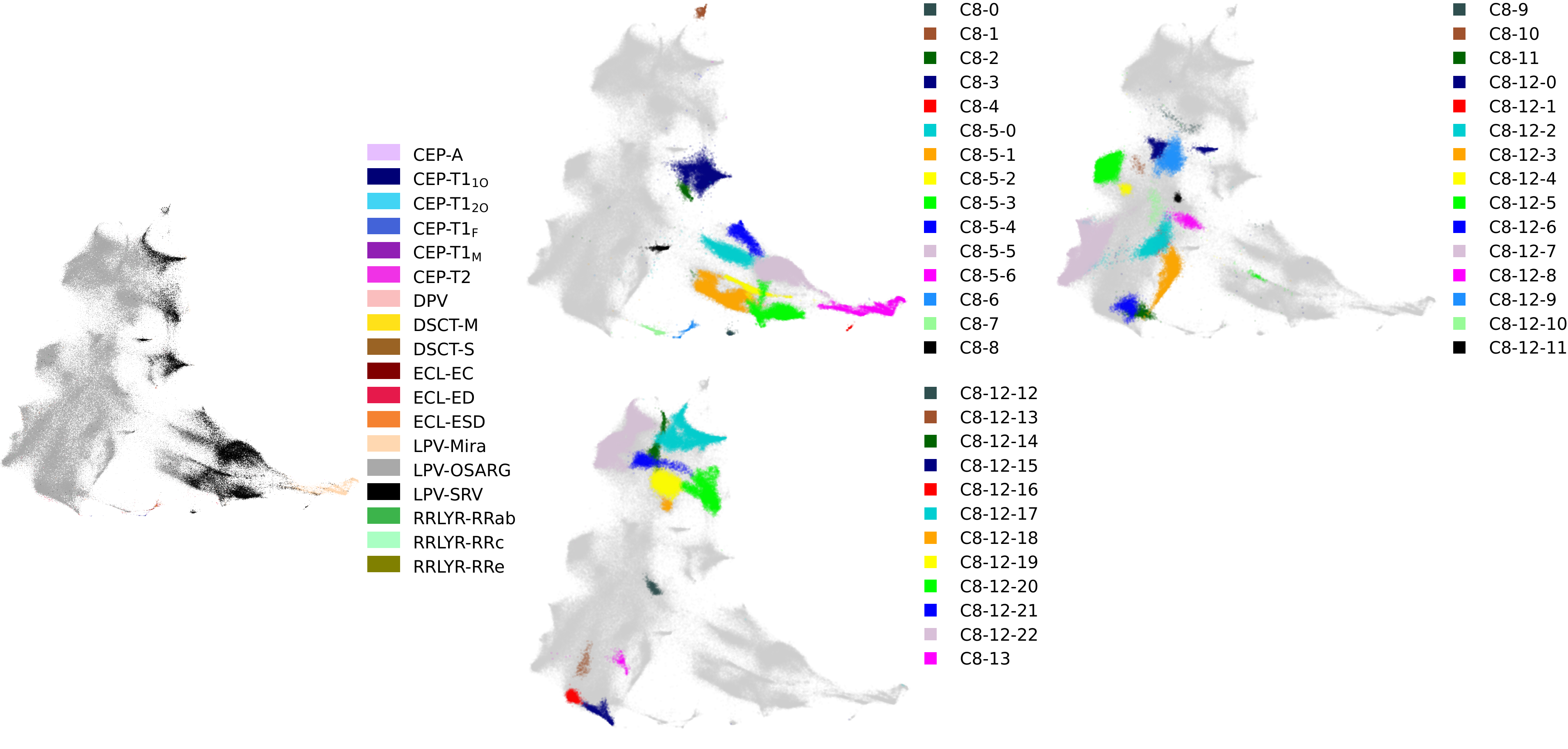}
    \caption{The left panel is a \textsc{umap} visualization of C8 from OCVS. The results on the \textsc{hdbscan} clustering done to cluster C8 from OCVS is shown in the three panel at the right. Gray points in the latter panels are drawn for reference purposes only, according to the full C8 data shown in the left panel.}
    \label{clust:ocvs4}
\end{figure*}

The LPV cluster C8 has several outliers and nested clusters that do not seem to match all the sub-classes completely. Figure~\ref{clust:ocvs4} presents the visualization of this cluster on the left and the clustering result in the three panels at its right.

In general, small clusters share interesting patterns; for example, cluster C8-4 has LPV light curves with trends in time, i.e., besides their large periods, they show dips, surges, or monotonic changes in time. Also, there are clusters highly correlated with the number of observations in the light curve. An example of this is C8-1, with a mixture of LPV-OSARGs and LPV-SRVs. The stars in this cluster have in common that a particular season was observed more frequently than other light curves of the same sub-classes.

Clusters C8-6 and C8-7 contain, in turn, LPVs, DPVs, long-period CEPs, and ECLs. The ECLs in these clusters are flagged as ellipsoidal systems by the OGLE team. It would be interesting to investigate in greater detail what properties may be shared with the ellipsoidals by the remaining variables in C8-6 and C8-7 that leads to them being grouped into such clusters. Moving on to larger clusters, we were able to find a pure cluster of LPV-Miras (C8-5-6 with 0.94) and plenty of LPV-OSARG clusters. As can be seen, many compact clusters surrounded by fuzzy ``noise'' exist in the LPV-OSARG structure at the left in Figure~\ref{clust:ocvs4}. Depending on the clustering parameters used, a significant number of these stars was dropped as noise by \textsc{hdbscan}; for this reason, the parameters that led to this result were carefully chosen. On the other hand, most of the LPV-SRV clusters suffer from severe contamination by LPV-OSARGs. This is not completely unexpected, considering that the different sub-classes of LPVs are not discreet, the boundaries between OSARGs and SRVs, and indeed between SRVs and Miras, being in fact arbitrary \citep{sosz-2013}. 

\begin{figure}
    \centering
    \includegraphics[width=1\columnwidth]{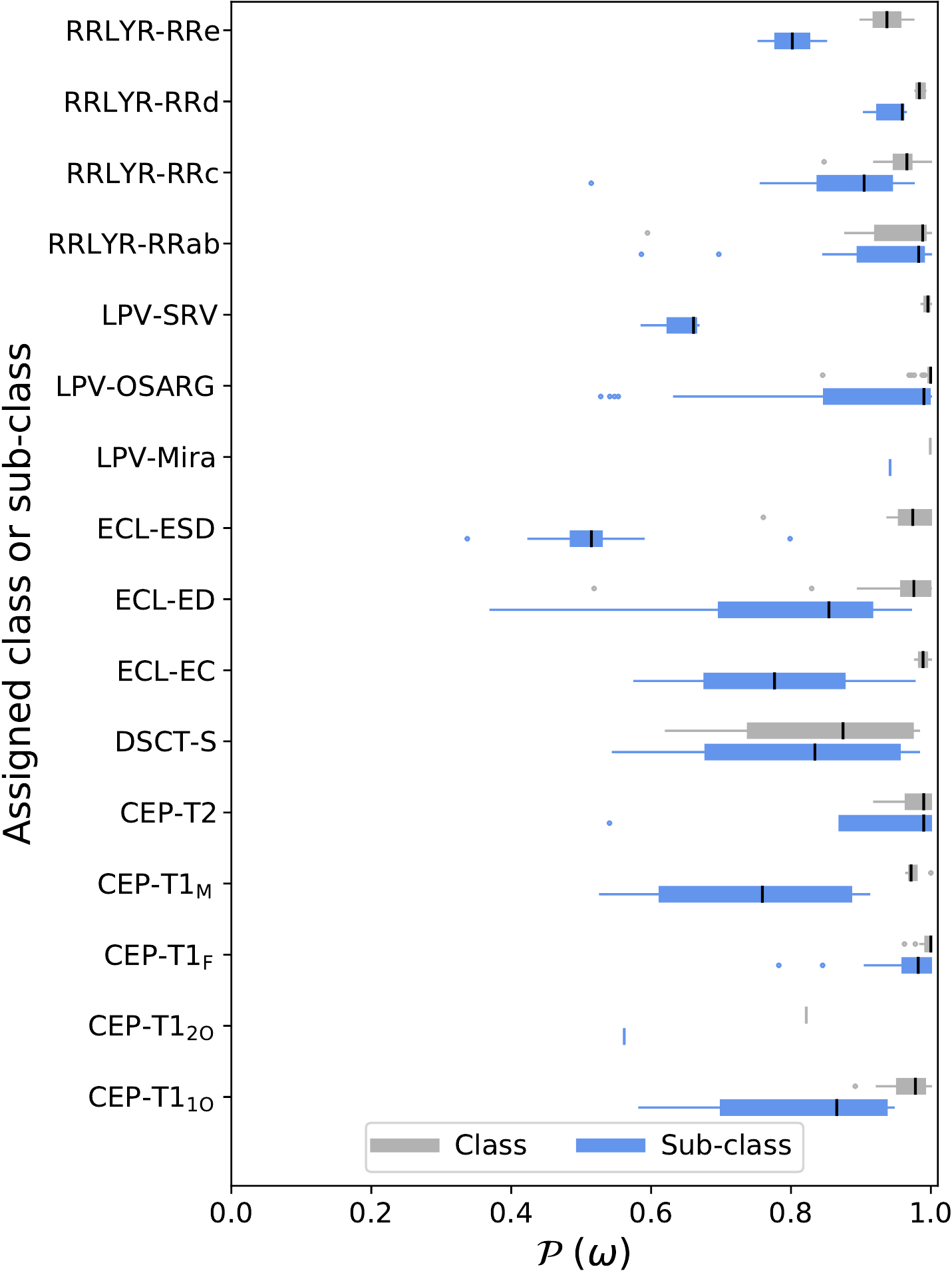}
    \caption{Cluster purity $\mathcal{P}(\omega)$ box-plots measured by class (gray) and sub-classes (light blue) assigned for OCVS. The black vertical lines indicate the median. If only one cluster is found, a vertical gray and a light blue line is drawn instead.}
    \label{purities:ocvs}
\end{figure}

In summary, we clustered 55\% of the data into 154 clusters. In Figure~\ref{purities:ocvs}, we show the results on the purity $\mathcal{P}(\omega)$ of the clusters found. In this figure, we see that there are sub-classes that are not completely separable. Table~\ref{tab:purities} confirms this too. In the CEP case, we see that we only can find pure clusters of CEP-T1$_{\rm F}$, CEP-T1$_{\rm 1O}$, and CEP-T2, but not of CEP-T1$_{\rm 2O}$ or CEP-T1$_{\rm M}$. Regarding the ECL classes, we are only able to find pure clusters of ECL-EC. The latter probably happened due to the lack of features to better characterize binary stars. However, the ECL clusters are highly pure, showing that other classes rarely contaminate them. With the LPV class, we found only one pure cluster of LPV-Miras. Also, we observed that the LPV-OSARG contaminates LPV-SRV clusters. Finally, the RRLYR class seems separable into sub-classes, with the exception of RRe stars, which are polluted by noisy or short-period variables such as DSCTs and some ECLs. 

\begin{table}
    \caption{Results on the purity, $\mathcal{P}(\Omega)$, measured respect to classes and sub-classes of the clusters corresponding to assigned sub-classes.}
    \label{tab:purities}
    \begin{tabular}{crcc}
    \hline
    \textbf{Catalog} & \textbf{Sub-class} & \textbf{$\mathcal{P}(\Omega)_{\text{Class}}$} & \textbf{$\mathcal{P}(\Omega)_{\text{Sub-class}}$} \\
    \hline
    \multirow{9}{*}{\textbf{CSSCVS}}       & CEP-T2     & 0.96 & 0.96 \\
                                  & DSCT       & 0.86 & \dotso \\
                                  & ECL-EA     & 0.99 & 0.74 \\
                                  & ECL-EW/EB  & 0.94 & 0.92 \\
                                  & LPV        & 0.97 & \dotso \\
                                  & ROT-RSCVn  & 0.61 & 0.61 \\
                                  & RRLYR-RRab & 0.91 & 0.74 \\
                                  & RRLYR-RRc  & 0.90 & 0.81 \\
                                  & RRLYR-RRd  & 0.91 & 0.58 \\
    \hline
    \multirow{5}{*}{\textbf{GDR2CVS}}      & CEP-T1     & 0.90 & 0.79 \\
                                  & DSCT/SXPHE & 0.83 & \dotso \\
                                  & MIRA/SRV   & 0.99 & \dotso \\
                                  & RRLYR-RRab & 0.97 & 0.95 \\
                                  & RRLYR-RRc  & 0.98 & 0.91 \\
    \hline                              
    \multirow{16}{*}{\textbf{OCVS}}        & CEP-T1$_{\rm 1O}$ & 0.98 & 0.89 \\
                                  & CEP-T1$_{\rm 2O}$ & 0.82 & 0.56 \\
                                  & CEP-T1$_{\rm F}$  & 0.99 & 0.96 \\
                                  & CEP-T1$_{\rm M}$  & 0.98 & 0.70 \\
                                  & CEP-T2     & 0.97 & 0.90 \\
                                  & DSCT-S     & 0.96 & 0.93 \\
                                  & ECL-EC     & 0.98 & 0.97 \\
                                  & ECL-ED     & 0.97 & 0.84 \\
                                  & ECL-ESD    & 0.92 & 0.48 \\
                                  & LPV-Mira   & 1.00 & 0.94 \\
                                  & LPV-OSARG  & 1.00 & 0.87 \\
                                  & LPV-SRV    & 1.00 & 0.66 \\
                                  & RRLYR-RRab & 0.98 & 0.97 \\
                                  & RRLYR-RRc  & 0.98 & 0.93 \\
                                  & RRLYR-RRd  & 0.99 & 0.94 \\
                                  & RRLYR-RRe  & 0.91 & 0.77 \\
    \hline                                  
    \end{tabular}
\end{table}

\section{Semi-supervised Classification}\label{sec:semisupervised}

Semi-supervised learning methods use the information from labeled data and their underlying distribution \citep{engelen2020}. This is accomplished by establishing that one or more of the following are true: the continuity assumption (same class samples are closer in feature space), the cluster assumption (same class samples are likely to share a discrete cluster), and the manifold assumption (there exists a lower-dimensional manifold in which most of the data lie). This translates to, for example, propagating labels inside the clusters (cluster assumption) or learning a metric from the data (continuity assumption).

Our semi-supervised method consists of a hierarchical procedure to classify classes of variable stars. It is designed to classify large groups of variable stars, transforming the data with Supervised \textsc{umap} DR and then performing classification with a Support Vector Machine \citep[SVM;][]{svm}. In this setting, two main assumptions are made: the cluster assumption, and manifold assumption. It requires less training set data than supervised methods, achieving comparable performance. Classification on large-scale surveys could benefit from it, especially when only a small training set is available from cross-matching or traditional variable star search methods. The complete process to classify a selected class (one at a time) is the following:
\begin{enumerate}
    \item \textbf{Feature scaling:} Perform a standard scaling of the data features.
    \item \textbf{Visualize the data:} Use \textsc{umap} to visualize the data. Color code the training set in the visualization to decide which class could be selected for classification. 
    \item \textbf{Prepare training set:} Convert the classes for binary classification, i.e., positive and negative classes. We only use training data available at this stage and reject the stars that were classified before.
    \item \textbf{Supervised \textsc{umap} DR training:} Train a \textsc{umap} supervised DR model using the available training set data. We chose to reduce the dimensionality to 20 dimensions with $\textit{n\_neighbours}=30$ to preserve the global structure. It is worth noting that tuning the \textit{target\_weight} parameter could be useful in case of having a low-quality or noisy training set.
    \item \textbf{Transform data:} Use \textsc{umap}'s learnt metric to transform the entire feature space (training set and unlabeled data). These data are now embedded into 20 dimensions and in two clusters (positive and negative classes), plus some noise in between.
    \item \textbf{SVM training:} Find the best Radial Basis Function (RBF) kernel SVM classifier (based on F$_1$-score) through a grid search trained with the 20-dimensional training set data in 5-fold cross-validation. The $C$ and $\gamma$ SVM hyperparameters are set between $10^{-3}$ and $10^{3}$ in log-scale steps.
    \item \textbf{SVM prediction:} Predict the classes of unlabeled data with the best SVM classifier.
\end{enumerate}

This method is applied hierarchically, classifying one class at a time. We use the training data available at each classification, not including the labeled data of previous classified classes. As a result, the negative class gets smaller after each classification. This allows the mitigation of the adverse effects of unbalanced splits. We strongly recommend using this method only to classify the data into classes that appear very well separated in the unsupervised \textsc{umap} visualization unless there is plenty of training set data for that particular class. Thus, the user will have to evaluate if a further split is safe or not based on the visualization.

We found that a non-linear RBF-SVM classifier achieves the highest performance compared with other classifiers such as Random Forest, Gaussian Process, Nearest Neighbors, and linear SVM. Indeed, the difference in performance is around 8\% on average. Regarding the hierarchical order of classification, we found that classifying large groups first (e.g., LPVs, ECLs, or pulsating variables) is better since finding classification margins for these groups is less complex, hence time complexity decreases in successive extractions (fewer data at each step), and it reduces the impact of very unbalanced splits. Note that it can be helpful (though not neccesary) to perform clustering analysis to have a more precise idea of the inherent hierarchical data structure.

In order to demonstrate the potential and weakness of our method, we bring it to its limits in the classification of our data, i.e., we will classify as many classes or sub-classes that score a reasonable F$_1$-score. Since the training set is small, we should expect variations in the classification results caused by the sampling, especially for minority classes. We confirmed this behavior in our tests, and we also found that the variance of the model itself, i.e., with a fixed training set sample, is negligible. Therefore, to account for how much exactly the training set sampling varies the classification, we gathered 25 different training set samples, all of them having the same size. We repeated the classification using these training sets following the corresponding hierarchical order to account for the final variance. All performance metrics and confusion matrices are measured from the test sets (i.e., including $\sim 95\%$ of the data) after all these stars have been classified. As a result, we find that performance metrics such as precision, recall, and F$_1$-score, have a standard deviation of around 0.004 for each catalog.

Additionally, we performed a test to see if the model generalizes well, using 15\%, 30\%, and 50\% of the data as training, finding that the F$_1$-score increases by an average of 0.03 for all our catalogs. The model quickly reaches very high levels of accuracy with a small amount of training data that marginally improves when adding additional data for training, suggesting that the model generalizes reliably.
Also, we noticed that the running time exponentially increases when using more training data, and the largest contribution comes from the cross-validation search of the best RBF-SVM classifier. Our method is designed for small training sets; however, in case one wishes to apply it to larger datasets, one option would be to use a linear-SVM or a random forest algorithm instead.

\subsection{Results on CSSCVS}

\begin{figure}
    \centering
    \includegraphics[width=0.5\columnwidth]{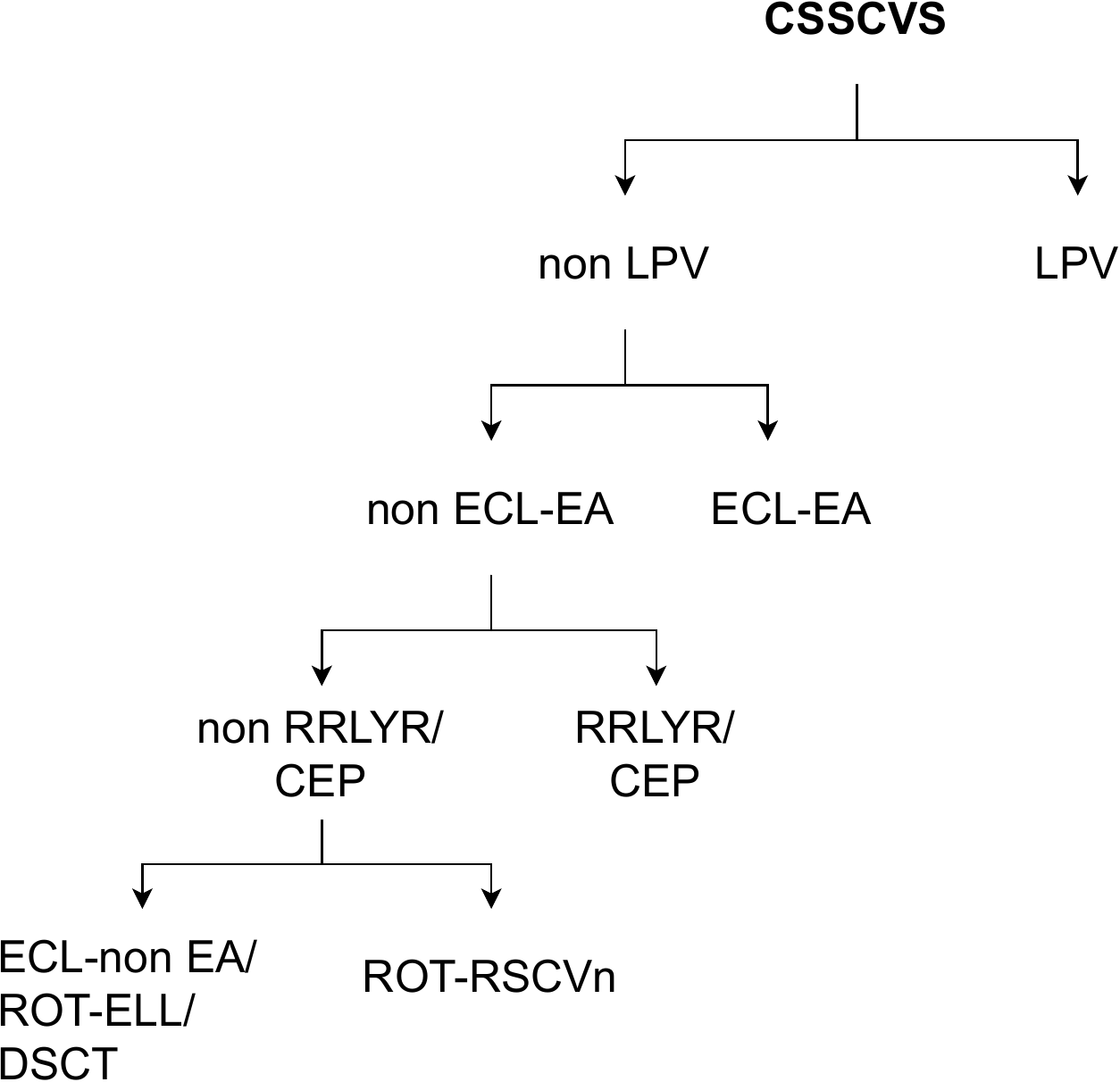}
    \caption{The hierarchical order of classification chosen for CSSCVS. Our semi-supervised model is trained and deployed for each parent node. The right child nodes represents the positive class.}
    \label{horder:csscvs}
\end{figure}

In this catalog, unbalanced sub-classes and a lower number of training variables contribute to only being able to classify large groups and having the lowest accuracy of our three catalogs. The first step was to study the visualization of this catalog in Figure~\ref{umap:csscvs}. In CSSCVS, the classification hierarchy was constructed based on the relative distance of clusters in the visualization while avoiding unbalanced splits. The hierarchical order of classification for this catalog is illustrated in Figure~\ref{horder:csscvs}. Each split in the tree represents a classification using our method (a training and a prediction phase), and each child at the right represents the positive class. As seen in this figure, we classify just a small number of classes or sub-classes available. This brings us to discuss two factors to consider when defining the hierarchical order: unbalanced classification and cluster merging. The unbalanced nature of this data complicates the classification of minority classes, thus setting a limit on the applicability of our method. Cluster merging is also a negative factor if a minority class spreads way beyond its primary cluster into another. An excellent example of both effects is the DSCT cluster. These stars lie in a cluster located at the bottom part of the visualization (shown in brown in Fig.~\ref{umap:csscvs}), but many spread smoothly into the ECL-EW/EB prime cluster. These two factors adversely combine: the DSCT cluster has not more than a hundred samples, and most of the DSCT stars are in the ECL-EW/EB cluster. As a result, the classification of this class is extremely poor (less than 0.5 accuracy), and so it should not be performed. Therefore, in the CSSCVS case, we do not provide further classification for the group containing mostly ECL-EW/EB (labeled as ECL-non EA/ROT-ELL/DSCT) stars, since various minority classes are almost completely mixed with ECL-EW/EB stars.

\begin{figure}
    \includegraphics[width=1\columnwidth]{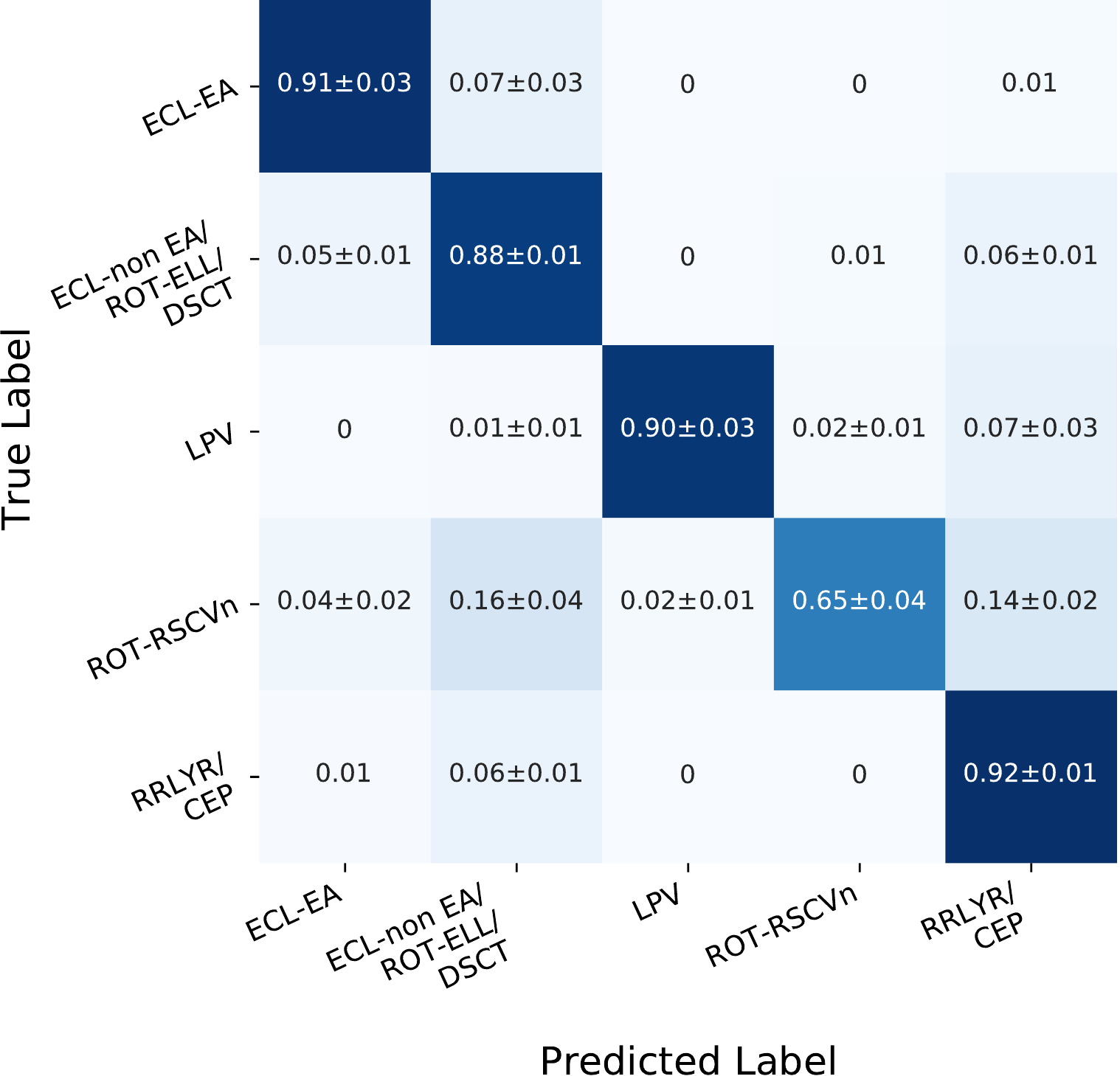}
    \caption{Confusion matrix obtained from our CSSCVS semi-supervised classification.}
    \label{cmatrix:csscvs}
\end{figure}

The final average performance metrics of the classification for this catalog are shown in Table~\ref{tab:performance}. As mentioned above, formal errors associated with the metrics in the table were not included since they were all near $0.004$. The confusion matrix in Figure~\ref{cmatrix:csscvs} includes standard errors associated to the averages values obtained from the 25 training set samples. However, averages of less than 0.01 are rounded as zeroes, and standard errors less than 0.01 are omitted for clarity.

We see that the ROT-RSCVn group does not achieve a good performance. This could be due to the low number of samples available for training (76) and the fact that these stars lie at different clusters in the embedding. Overall, in CSSCVS, we achieve a classification performance of around 89\%. 

\begin{table}
    \caption{Performance metrics of our semi-supervised classification.}
    \label{tab:performance}
    \begin{tabular}{cccc}
    \hline
    \textbf{Catalog} & \textbf{Precision} & \textbf{Recall} & \textbf{F$_1$-score} \\
    \hline
    CSSCVS   & 0.90 & 0.89 & 0.89 \\
    GDR2CVS  & 0.94 & 0.92 & 0.93 \\
    OCVS     & 0.93 & 0.91 & 0.92 \\   
    \hline
    \end{tabular}
\end{table}

\subsection{Results on GDR2CVS}

\begin{figure}
    \centering
    \includegraphics[width=0.6\columnwidth]{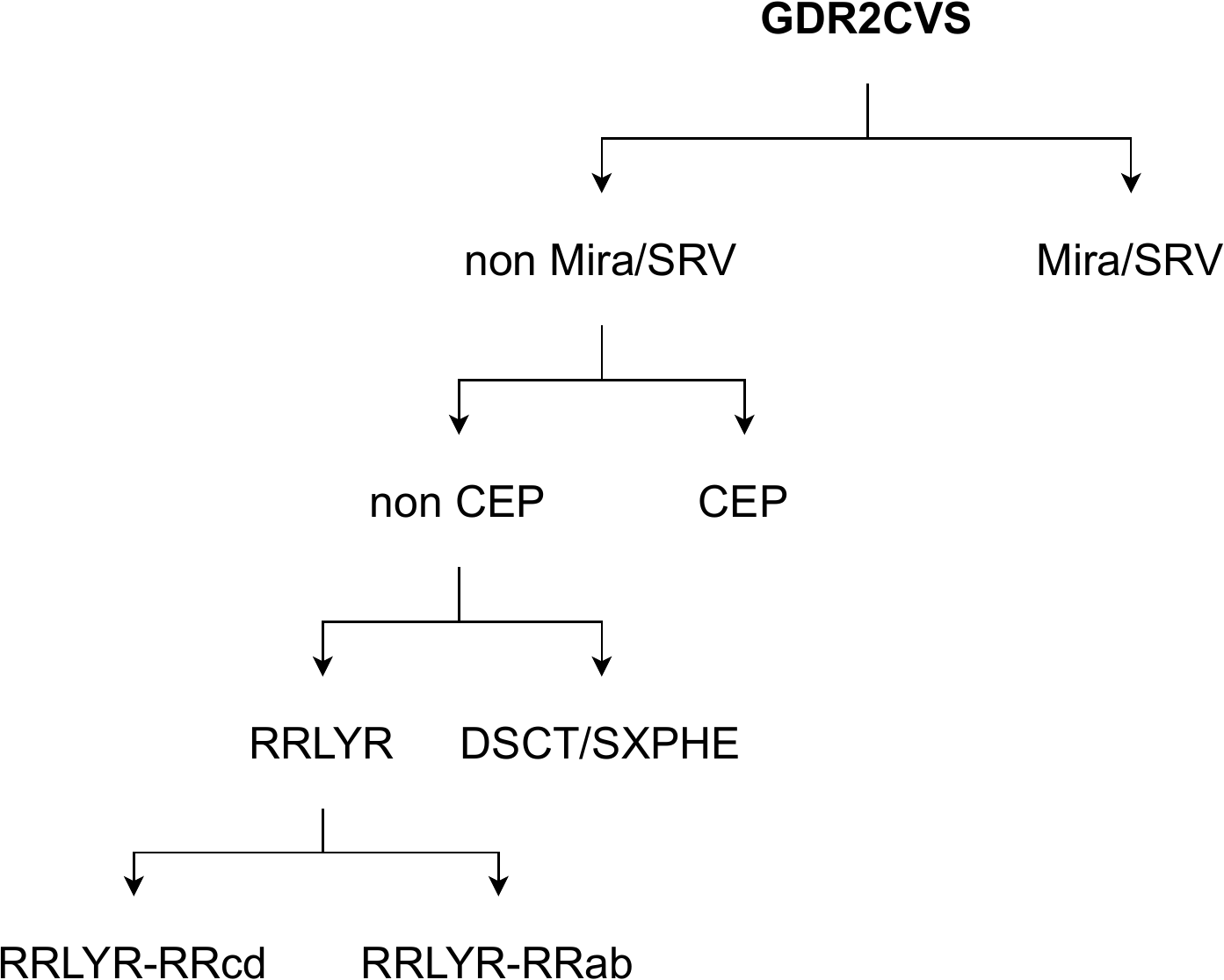}
    \caption{As in Figure~\ref{horder:csscvs}, but for GDR2CVS.}
    \label{horder:gdr2cvs}
\end{figure}

In GDR2CVS, the choice of the order of extraction of classes is straightforward. There are fewer sub-classes in this catalog, and most of them lie in one or more rather outlined clusters. Accordingly, we extracted these sub-classes as shown in Figure~\ref{horder:gdr2cvs}. 

\begin{figure}
    \includegraphics[width=1\columnwidth]{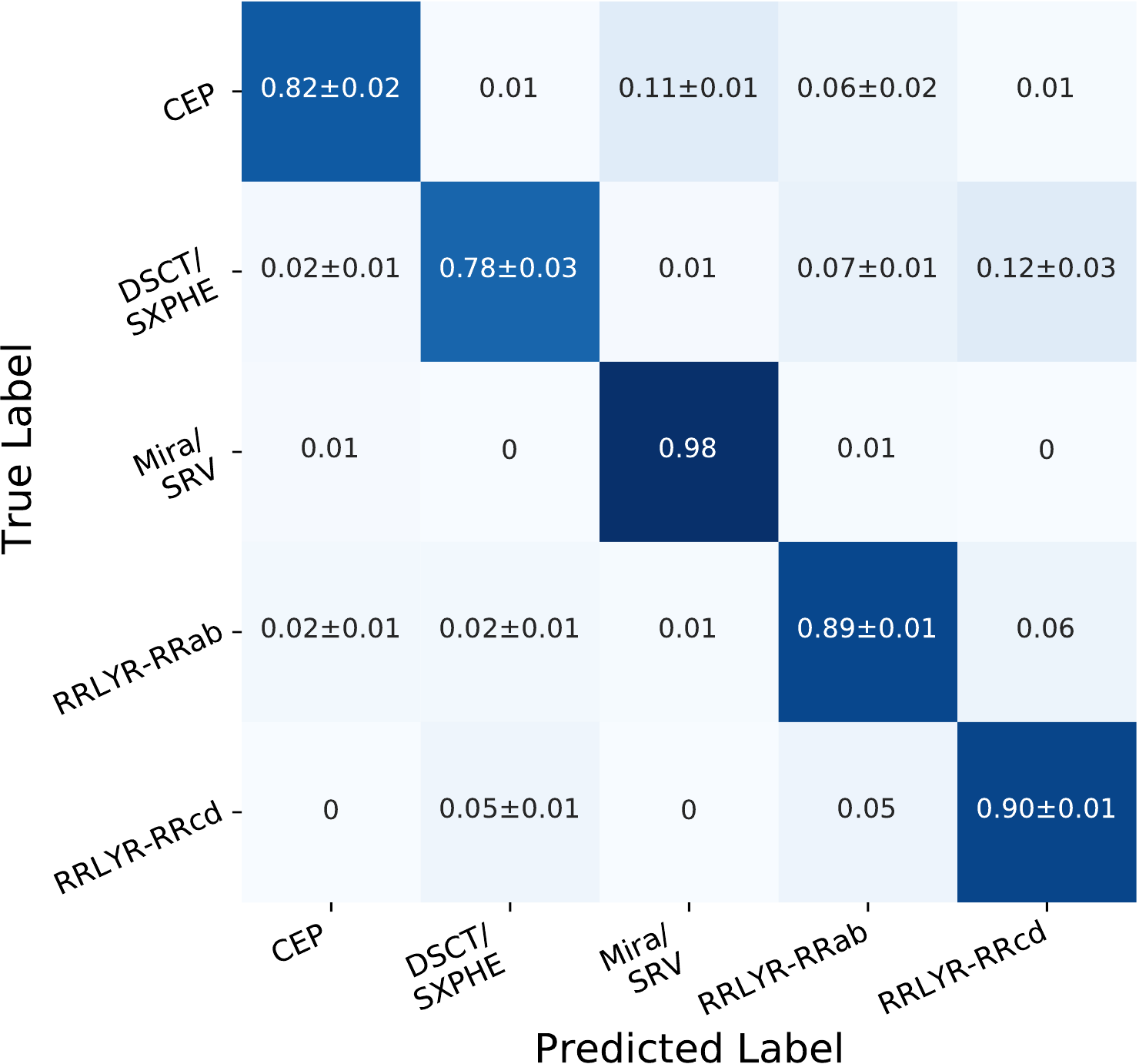}
    \caption{As in Figure~\ref{cmatrix:csscvs}, but for GDR2CVS.}
    \label{cmatrix:grd2cvs}
\end{figure}

The confusion matrix for this classification is portrayed in Figure~\ref{cmatrix:grd2cvs}. In this matrix, we notice that there is significant cross-contamination between the classes. This is the case of DSCT/SXPHE, which has a large amount of RRc contaminants. Moreover, it is interesting that 11\% of CEP are confounded in the Mira/SRV cluster. This group contains mostly CEP-T2 (possibly RVTau), and they populate the right edge of the largest LPV cluster in the visualization. Finally, according to Table~\ref{tab:performance}, for this catalog, we achieve an average performance of about 93\%. It is worth noting that this high F$_1$-score is in great measure due to the numerous three largest classes, Mira/SRV, RRLYR-RRab, and RRLYR-RRcd.

\subsection{Results on OCVS}

\begin{figure}
    \centering
    \includegraphics[width=0.95\columnwidth]{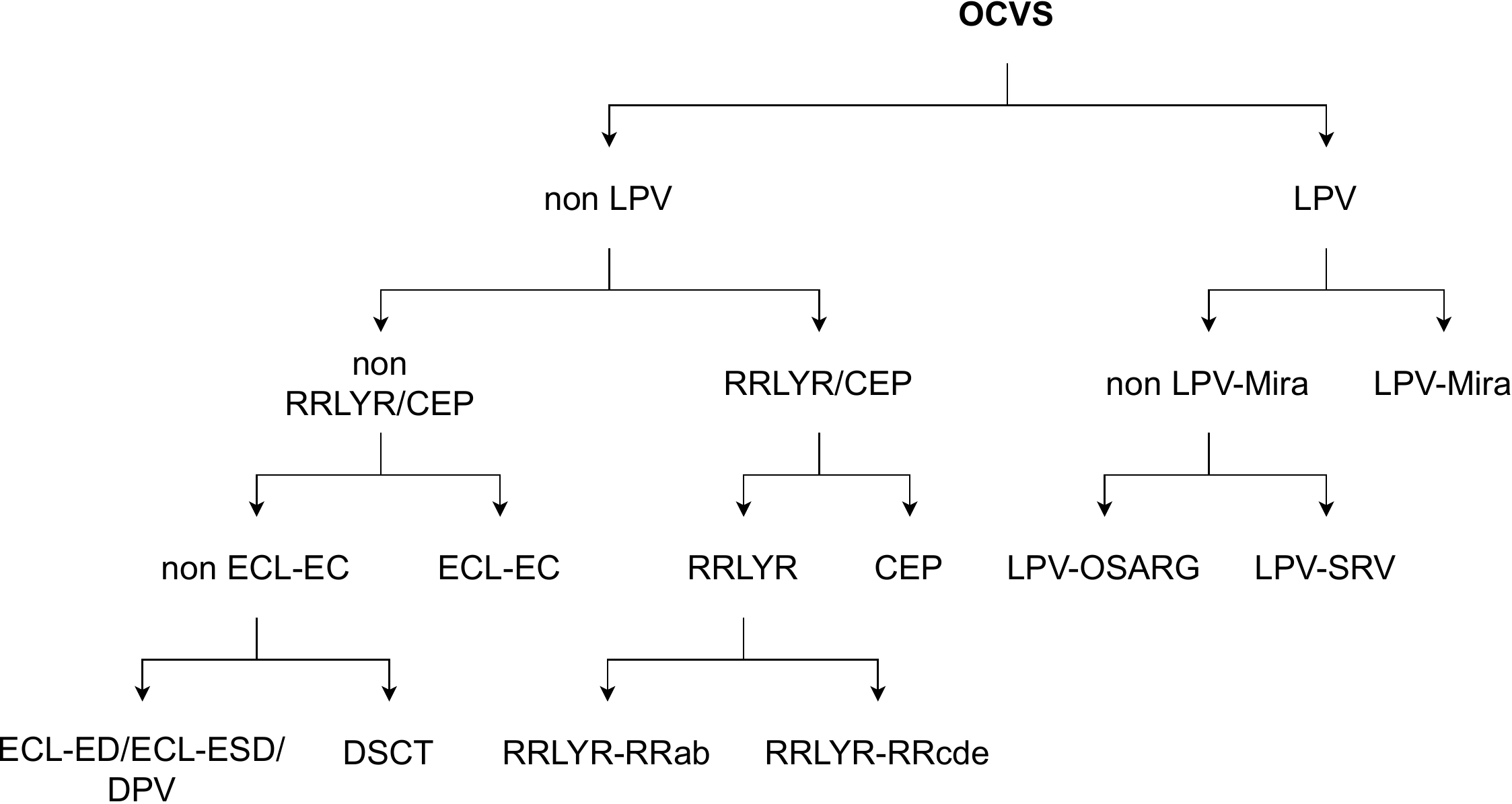}
    \caption{As in Figure~\ref{horder:csscvs}, but for OCVS.}
    \label{horder:ocvs}
\end{figure}

\begin{figure*}
    \centering
    \includegraphics[width=1.5\columnwidth]{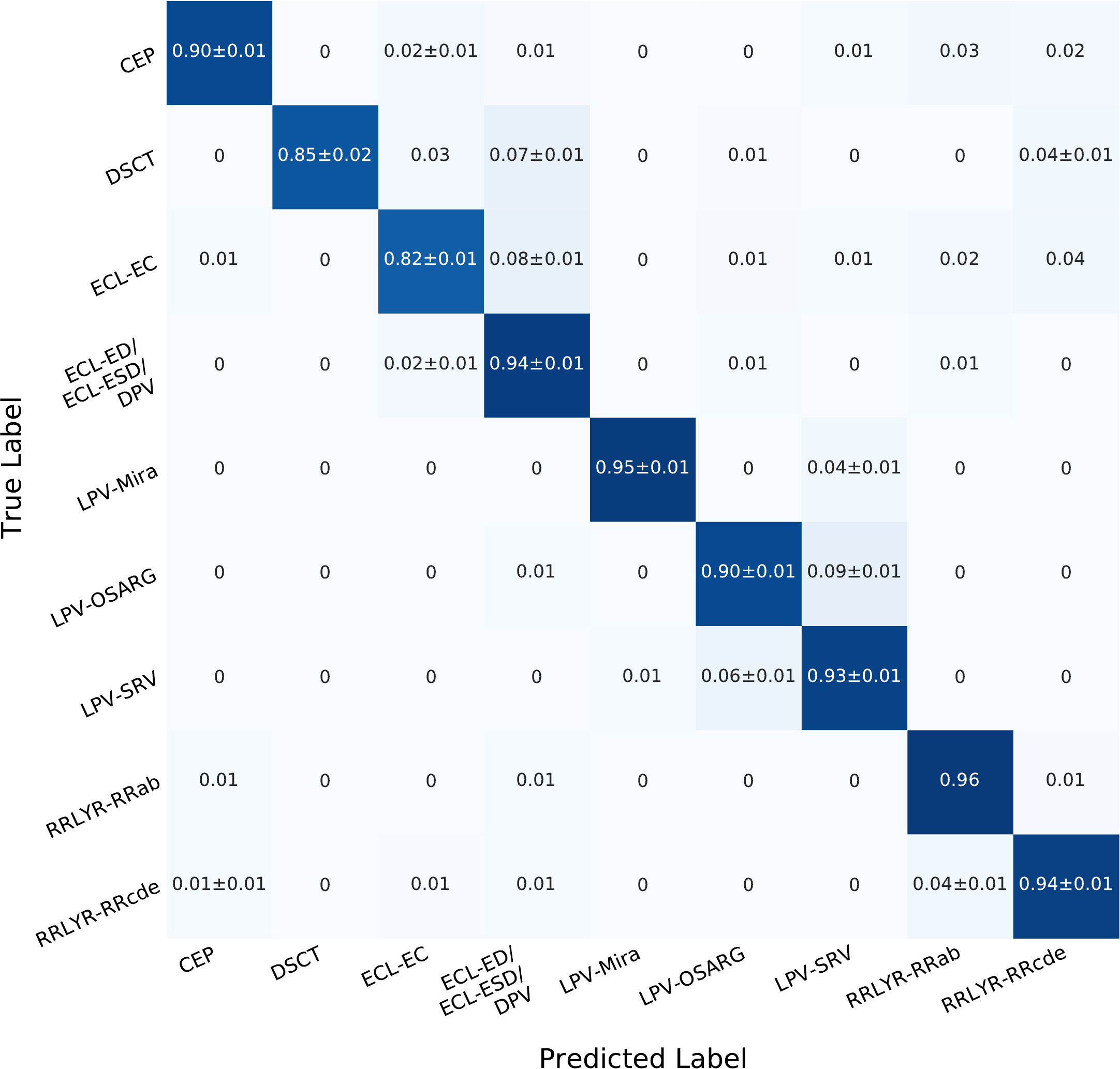}
    \caption{As in Figure~\ref{cmatrix:csscvs}, but for OCVS.}
    \label{cmatrix:ocvs}
\end{figure*}

OCVS has high-quality data, so we were able to classify many sub-classes of variable stars. This is confirmed by its visualization, where we clearly observe how these sub-classes form clusters. The order of extraction is done according to Figure~\ref{horder:ocvs}, always trying to avoid unbalanced splits. The average metrics are shown in Table~\ref{tab:performance} and the average confusion matrix is shown in Figure~\ref{cmatrix:ocvs}.

We observe the usual cross-contamination in this classification, namely LPV-SRV \& LPV-OSARG, LPV-Mira \& LPV-SRV, CEP \& RRLYR-RRab and DSCT \& RRLYR-RRcde. As seen in the other catalogs, we notice that classification performance on minority classes is limited. However, for the rest of the variables, the classification performance is near or over 90\%. Finally, from Table~\ref{tab:performance} we see that the classification of this catalog achieves an average performance of about 92\%.

\section{Discussion}\label{sec:discussion}

Research on variable star classification in large-scale surveys is expanding rapidly as new massive data is released. We expect a substantial positive impact in astrophysics from these studies as they reach unprecedented levels of accuracy. However, the classifiers implemented in these studies heavily rely on a high-quality training set, i.e., thousands of samples per sub-class or more. Building a training set is expensive in terms of time and complexity, which puts a serious bound to the progress of supervised learning methods. In this paper, we tackled this issue, offering novel solutions for variable star classification and clustering. First, we engineered new features based on the periodogram to complement traditional variable star features. Second, we offered an alternative semi-supervised approach to use small training sets to classify the main classes of variable stars, achieving good levels of accuracy. Third, we put forward a fully unsupervised clustering procedure to explore, discover and group variable stars. DR with the \textsc{umap} algorithm is one of the keystones of our methods, significantly improving the latter's final performance and giving valuable insights into the data through visualizations.

Our semi-supervised method for the classification of variable stars is tested using three different catalogs and trained with only 5\% of the catalog's stars, achieving around 90\% of accuracy. This method requires the user to decide how the classes are classified based on \textsc{umap} visualizations or other prior knowledge about the data. Depending on the data and the size of the available training set, it is even suitable for classifying sub-classes. However, in the case of a very small training set, we recommend the user classify only the classes or large groups of variable stars. There are many ways to improve upon our method, such as oversampling minority classes with the synthetic minority oversampling technique (SMOTE, \citealt{chawla2010}) to improve its accuracy when the training set is very unbalanced. 

Finally, a comprehensive unsupervised clustering analysis procedure was devised to explore our variable star catalogs with the help of \textsc{umap} visualizations. We found clusters with very high purity concerning classes and some sub-classes. This study demonstrates that this type of exploratory analysis can be very reliable for finding classes or sub-classes of variable stars and potentially be very effective for serendipitous discovery. We learned that some classes are not entirely separable into their respective sub-classes with our current light curve features, and some minority classes could be missed. There are options to mitigate these issues, such as re-clustering the noise assigned by \textsc{hdbscan} or dividing the clustering analysis into more hierarchical levels. Moreover, it may be more appropriate to use specialized features when dealing with a cluster of a known class, e.g., to derive eclipse parameters for ECL (to distinguish between ECL-ESD and ECL-ED) or to use entropy-based periodograms to better characterize semi-periodic variables (e.g., LPV-SRVs or LPV-OSARGs). Doing clustering analysis on large-scale surveys can be extremely useful to describe the dataset itself, including its particular intricacies, and gain a broad understanding of the variable star zoo concealed in the data.

\section*{Acknowledgements}
This work has received funded by the National Agency for Research and Development (ANID), PFCHA/DOCTORADO NACIONAL/2015-21151132. Support for R.P. and M.C. is also provided by ANID's Millennium Science Initiative through grant ICN12\textunderscore 12009, awarded to the Millennium Institute of Astrophysics (MAS); by Proyecto Basal FB210003 and AFB-170002; and by FONDECYT grant \#1171273. 

\section*{Data Availability}
The data generated in this paper is available in the first author's GitHub page, at \url{https://github.com/rdpantoja/clustering_variable_stars_2022}.



\bibliographystyle{mnras}
\bibliography{biblio} 

\providecommand{\noopsort}[1]{}
\begin{thebibliography}{}
\makeatletter
\relax
\def\mn@urlcharsother{\let\do\@makeother \do\$\do\&\do\#\do\^\do\_\do\%\do\~}
\def\mn@doi{\begingroup\mn@urlcharsother \@ifnextchar [ {\mn@doi@}
  {\mn@doi@[]}}
\def\mn@doi@[#1]#2{\def\@tempa{#1}\ifx\@tempa\@empty \href
  {http://dx.doi.org/#2} {doi:#2}\else \href {http://dx.doi.org/#2} {#1}\fi
  \endgroup}
\def\mn@eprint#1#2{\mn@eprint@#1:#2::\@nil}
\def\mn@eprint@arXiv#1{\href {http://arxiv.org/abs/#1} {{\tt arXiv:#1}}}
\def\mn@eprint@dblp#1{\href {http://dblp.uni-trier.de/rec/bibtex/#1.xml}
  {dblp:#1}}
\def\mn@eprint@#1:#2:#3:#4\@nil{\def\@tempa {#1}\def\@tempb {#2}\def\@tempc
  {#3}\ifx \@tempc \@empty \let \@tempc \@tempb \let \@tempb \@tempa \fi \ifx
  \@tempb \@empty \def\@tempb {arXiv}\fi \@ifundefined
  {mn@eprint@\@tempb}{\@tempb:\@tempc}{\expandafter \expandafter \csname
  mn@eprint@\@tempb\endcsname \expandafter{\@tempc}}}

\bibitem[\protect\citeauthoryear{{Aerts}}{{Aerts}}{2021}]{aerts-2021}
{Aerts} C.,  2021, \mn@doi [Reviews of Modern Physics]
  {10.1103/RevModPhys.93.015001}, \href
  {https://ui.adsabs.harvard.edu/abs/2021RvMP...93a5001A} {93, 015001}

\bibitem[\protect\citeauthoryear{{Aerts}, {Christensen-Dalsgaard}  \&
  {Kurtz}}{{Aerts} et~al.}{2010}]{aerts2010}
{Aerts} C.,  {Christensen-Dalsgaard} J.,   {Kurtz} D.~W.,  2010,
  {Asteroseismology}.
Springer, Dordrecht

\bibitem[\protect\citeauthoryear{{Aggarwal} \& {Reddy}}{{Aggarwal} \&
  {Reddy}}{2013}]{aggarwal2013}
{Aggarwal} C.,  {Reddy} C.,  2013, {Data Clustering: Algorithms and
  Applications}.
Chapman and Hall/CRC, \mn@doi{https://doi.org/10.1201/9781315373515}

\bibitem[\protect\citeauthoryear{{Aguirre}, {Pichara}  \& {Becker}}{{Aguirre}
  et~al.}{2019}]{aguirre2019}
{Aguirre} C.,  {Pichara} K.,   {Becker} I.,  2019, \mn@doi [\mnras]
  {10.1093/mnras/sty2836}, \href
  {https://ui.adsabs.harvard.edu/abs/2019MNRAS.482.5078A} {482, 5078}

\bibitem[\protect\citeauthoryear{{Angeloni} et~al.,}{{Angeloni}
  et~al.}{2014}]{angeloni2014}
{Angeloni} R.,  et~al., 2014, \mn@doi [\aap] {10.1051/0004-6361/201423904},
  \href {https://ui.adsabs.harvard.edu/abs/2014A&A...567A.100A} {567, A100}

\bibitem[\protect\citeauthoryear{{Astropy Collaboration} et~al.,}{{Astropy
  Collaboration} et~al.}{2013}]{astropy:2013}
{Astropy Collaboration} et~al., 2013, \mn@doi [\aap]
  {10.1051/0004-6361/201322068}, \href
  {http://adsabs.harvard.edu/abs/2013A%26A...558A..33A} {558, A33}

\bibitem[\protect\citeauthoryear{{Astropy Collaboration} et~al.,}{{Astropy
  Collaboration} et~al.}{2018}]{astropy:2018}
{Astropy Collaboration} et~al., 2018, \mn@doi [\aj] {10.3847/1538-3881/aabc4f},
  \href {https://ui.adsabs.harvard.edu/abs/2018AJ....156..123A} {156, 123}

\bibitem[\protect\citeauthoryear{{Becker}, {Pichara}, {Catelan}, {Protopapas},
  {Aguirre}  \& {Nikzat}}{{Becker} et~al.}{2020}]{becker2020}
{Becker} I.,  {Pichara} K.,  {Catelan} M.,  {Protopapas} P.,  {Aguirre} C.,
  {Nikzat} F.,  2020, \mn@doi [\mnras] {10.1093/mnras/staa350}, \href
  {https://ui.adsabs.harvard.edu/abs/2020MNRAS.493.2981B} {493, 2981}

\bibitem[\protect\citeauthoryear{{Benavente}, {Protopapas}  \&
  {Pichara}}{{Benavente} et~al.}{2017}]{benavente2017}
{Benavente} P.,  {Protopapas} P.,   {Pichara} K.,  2017, \mn@doi [\apj]
  {10.3847/1538-4357/aa7f2d}, \href
  {https://ui.adsabs.harvard.edu/abs/2017ApJ...845..147B} {845, 147}

\bibitem[\protect\citeauthoryear{{Brink}, {Richards}, {Poznanski}, {Bloom},
  {Rice}, {Negahban}  \& {Wainwright}}{{Brink} et~al.}{2013}]{brink2013}
{Brink} H.,  {Richards} J.~W.,  {Poznanski} D.,  {Bloom} J.~S.,  {Rice} J.,
  {Negahban} S.,   {Wainwright} M.,  2013, \mn@doi [\mnras]
  {10.1093/mnras/stt1306}, \href
  {https://ui.adsabs.harvard.edu/abs/2013MNRAS.435.1047B} {435, 1047}

\bibitem[\protect\citeauthoryear{{Brys}, {Hubert}  \& {Struyf}}{{Brys}
  et~al.}{2004}]{brys2004}
{Brys} G.,  {Hubert} M.,   {Struyf} A.,  2004, \mn@doi [Journal of
  Computational and Graphical Statistics] {10.1198/106186004X12632}, 13, 996

\bibitem[\protect\citeauthoryear{{Brys}, {Hubert}  \& {Struyf}}{{Brys}
  et~al.}{2006}]{brys2006}
{Brys} G.,  {Hubert} M.,   {Struyf} A.,  2006, \mn@doi [Computational
  Statistics \& Data Analysis] {10.1016/j.csda.2004.09.012}, 50, 733

\bibitem[\protect\citeauthoryear{{Butler} \& {Bloom}}{{Butler} \&
  {Bloom}}{2011}]{butler2011}
{Butler} N.~R.,  {Bloom} J.~S.,  2011, \mn@doi [\aj]
  {10.1088/0004-6256/141/3/93}, \href
  {https://ui.adsabs.harvard.edu/abs/2011AJ....141...93B} {141, 93}

\bibitem[\protect\citeauthoryear{{Campello}, {Moulavi}  \& {Sander}}{{Campello}
  et~al.}{2013}]{campello2013}
{Campello} R.,  {Moulavi} D.,   {Sander} J.,  2013, in Advances in Knowledge
  Discovery and Data Mining. Springer Berlin Heidelberg, Berlin, Heidelberg, pp
  160--172, \mn@doi{10.1007/978-3-642-37456-2_14}

\bibitem[\protect\citeauthoryear{{Castro}, {Protopapas}  \& {Pichara}}{{Castro}
  et~al.}{2018}]{castro2018}
{Castro} N.,  {Protopapas} P.,   {Pichara} K.,  2018, \mn@doi [\aj]
  {10.3847/1538-3881/aa9ab8}, \href
  {https://ui.adsabs.harvard.edu/abs/2018AJ....155...16C} {155, 16}

\bibitem[\protect\citeauthoryear{{Catelan} \& {Smith}}{{Catelan} \&
  {Smith}}{2015}]{catelan2015}
{Catelan} M.,  {Smith} H.~A.,  2015, {Pulsating Stars}.
Wiley-VCH, Weinheim

\bibitem[\protect\citeauthoryear{{Chapelle}, {Sch\"olkopf}  \&
  {Zien}}{{Chapelle} et~al.}{2006}]{chapelle2006}
{Chapelle} O.,  {Sch\"olkopf} B.,   {Zien} A.,  2006, Semi-Supervised Learning,
  1st edn.
The MIT Press, \mn@doi{10.7551/mitpress/9780262033589.001.0001}

\bibitem[\protect\citeauthoryear{{Chawla}, {Bowyer}, {Hall}  \&
  {Kegelmeyer}}{{Chawla} et~al.}{2011}]{chawla2010}
{Chawla} N.~V.,  {Bowyer} K.~W.,  {Hall} L.~O.,   {Kegelmeyer} W.~P.,  2011,
  arXiv e-prints, \href {https://ui.adsabs.harvard.edu/abs/2011arXiv1106.1813C}
  {p. arXiv:1106.1813}

\bibitem[\protect\citeauthoryear{{Christensen-Dalsgaard}}{{Christensen-Dalsgaard}}{2002}]{CD-2002}
{Christensen-Dalsgaard} J.,  2002, \mn@doi [Reviews of Modern Physics]
  {10.1103/RevModPhys.74.1073}, \href
  {https://ui.adsabs.harvard.edu/abs/2002RvMP...74.1073C} {74, 1073}

\bibitem[\protect\citeauthoryear{{Christensen-Dalsgaard}}{{Christensen-Dalsgaard}}{2021}]{CD-2021}
{Christensen-Dalsgaard} J.,  2021, \mn@doi [Living Reviews in Solar Physics]
  {10.1007/s41116-020-00028-3}, \href
  {https://ui.adsabs.harvard.edu/abs/2021LRSP...18....2C} {18, 2}

\bibitem[\protect\citeauthoryear{{Cioni} et~al.,}{{Cioni}
  et~al.}{2011}]{Cioni-2011}
{Cioni} M. R.~L.,  et~al., 2011, \mn@doi [\aap] {10.1051/0004-6361/201016137},
  \href {https://ui.adsabs.harvard.edu/abs/2011A&A...527A.116C} {527, A116}

\bibitem[\protect\citeauthoryear{{Cook} et~al.,}{{Cook} et~al.}{1997}]{macho}
{Cook} K.~H.,  et~al., 1997, in {Ferlet} R.,  {Maillard} J.-P.,   {Raban} B.,
  eds, Variables Stars and the Astrophysical Returns of the Microlensing
  Surveys. p.~17

\bibitem[\protect\citeauthoryear{Cortes \& Vapnik}{Cortes \&
  Vapnik}{1995}]{svm}
Cortes C.,  Vapnik V.,  1995, \mn@doi [Machine learning] {10.1007/BF00994018},
  20, 273

\bibitem[\protect\citeauthoryear{{Cox}}{{Cox}}{1980}]{cox1980}
{Cox} J.~P.,  1980, Theory of stellar pulsation.
Princeton University Press, Princeton, New Jersey

\bibitem[\protect\citeauthoryear{{Deb} \& {Singh}}{{Deb} \&
  {Singh}}{2009}]{deb2009}
{Deb} S.,  {Singh} H.~P.,  2009, \mn@doi [\aap] {10.1051/0004-6361/200912851},
  \href {https://ui.adsabs.harvard.edu/abs/2009A&A...507.1729D} {507, 1729}

\bibitem[\protect\citeauthoryear{{Debosscher}, {Sarro}, {Aerts}, {Cuypers},
  {Vandenbussche}, {Garrido}  \& {Solano}}{{Debosscher}
  et~al.}{2007}]{debosscher2007}
{Debosscher} J.,  {Sarro} L.~M.,  {Aerts} C.,  {Cuypers} J.,  {Vandenbussche}
  B.,  {Garrido} R.,   {Solano} E.,  2007, \mn@doi [\aap]
  {10.1051/0004-6361:20077638}, \href
  {https://ui.adsabs.harvard.edu/abs/2007A&A...475.1159D} {475, 1159}

\bibitem[\protect\citeauthoryear{{Drake} et~al.,}{{Drake}
  et~al.}{2009}]{drake2009}
{Drake} A.~J.,  et~al., 2009, \mn@doi [\apj] {10.1088/0004-637X/696/1/870},
  \href {https://ui.adsabs.harvard.edu/abs/2009ApJ...696..870D} {696, 870}

\bibitem[\protect\citeauthoryear{{Drake} et~al.,}{{Drake}
  et~al.}{2014}]{drake2014}
{Drake} A.~J.,  et~al., 2014, \apjs, \href
  {http://adsabs.harvard.edu/abs/2014ApJS..213....9D} {213, 9}

\bibitem[\protect\citeauthoryear{{Drake} et~al.,}{{Drake}
  et~al.}{2017}]{drake2017}
{Drake} A.~J.,  et~al., 2017, \mnras, \href
  {http://adsabs.harvard.edu/abs/2017MNRAS.469.3688D} {469, 3688}

\bibitem[\protect\citeauthoryear{{Dubath} et~al.,}{{Dubath}
  et~al.}{2011}]{dubath2011}
{Dubath} P.,  et~al., 2011, \mn@doi [\mnras]
  {10.1111/j.1365-2966.2011.18575.x}, \href
  {https://ui.adsabs.harvard.edu/abs/2011MNRAS.414.2602D} {414, 2602}

\bibitem[\protect\citeauthoryear{Eddington}{Eddington}{1918}]{eddington1918}
Eddington A.~S.,  1918, \mn@doi [\mnras] {10.1093/mnras/79.1.2}, 79, 2

\bibitem[\protect\citeauthoryear{{Elorrieta} et~al.,}{{Elorrieta}
  et~al.}{2016}]{elorrieta2016}
{Elorrieta} F.,  et~al., 2016, \mn@doi [\aap] {10.1051/0004-6361/201628700},
  \href {https://ui.adsabs.harvard.edu/abs/2016A&A...595A..82E} {595, A82}

\bibitem[\protect\citeauthoryear{{Ester}, {Kriegel}, {Sander}  \& {Xu}}{{Ester}
  et~al.}{1996}]{ester1996}
{Ester} M.,  {Kriegel} H.,  {Sander} J.,   {Xu} X.,  1996, in Proceedings of
  the Second International Conference on Knowledge Discovery and Data Mining.
  AAAI Press, pp 226--231, \url
  {http://dl.acm.org/citation.cfm?id=3001460.3001507}

\bibitem[\protect\citeauthoryear{{Eyer}, {S{\"u}veges}, {De Ridder}, {Regibo},
  {Mowlavi}, {Holl}, {Rimoldini}  \& {Bouchy}}{{Eyer} et~al.}{2019}]{eyer2019}
{Eyer} L.,  {S{\"u}veges} M.,  {De Ridder} J.,  {Regibo} S.,  {Mowlavi} N.,
  {Holl} B.,  {Rimoldini} L.,   {Bouchy} F.,  2019, \mn@doi [\pasp]
  {10.1088/1538-3873/ab2511}, \href
  {https://ui.adsabs.harvard.edu/abs/2019PASP..131h8001E} {131, 088001}

\bibitem[\protect\citeauthoryear{{Ferreira Lopes} \& {Cross}}{{Ferreira Lopes}
  \& {Cross}}{2017}]{ferreiralopes2017}
{Ferreira Lopes} C.~E.,  {Cross} N.~J.~G.,  2017, \mn@doi [\aap]
  {10.1051/0004-6361/201630109}, \href
  {https://ui.adsabs.harvard.edu/abs/2017A&A...604A.121F} {604, A121}

\bibitem[\protect\citeauthoryear{{Fisher}}{{Fisher}}{1936}]{fisher1936}
{Fisher} R.~A.,  1936, \mn@doi [Annals of Eugenics]
  {https://doi.org/10.1111/j.1469-1809.1936.tb02137.x}, 7, 179

\bibitem[\protect\citeauthoryear{{Gaia Collaboration} et~al.,}{{Gaia
  Collaboration} et~al.}{2016}]{gaiamission}
{Gaia Collaboration} et~al., 2016, \mn@doi [\aap]
  {10.1051/0004-6361/201629272}, \href
  {https://ui.adsabs.harvard.edu/abs/2016A&A...595A...1G} {595, A1}

\bibitem[\protect\citeauthoryear{{Gaia Collaboration} et~al.,}{{Gaia
  Collaboration} et~al.}{2018}]{gaiadr2}
{Gaia Collaboration} et~al., 2018, \mn@doi [\aap]
  {10.1051/0004-6361/201833051}, \href
  {https://ui.adsabs.harvard.edu/abs/2018A&A...616A...1G} {616, A1}

\bibitem[\protect\citeauthoryear{{Graczyk} et~al.,}{{Graczyk}
  et~al.}{2011}]{graczyk2011}
{Graczyk} D.,  et~al., 2011, \actaa, \href
  {https://ui.adsabs.harvard.edu/abs/2011AcA....61..103G} {61, 103}

\bibitem[\protect\citeauthoryear{{Hassan}, {Mirabal}, {Contreras}  \&
  {Oya}}{{Hassan} et~al.}{2013}]{hassan2013}
{Hassan} T.,  {Mirabal} N.,  {Contreras} J.~L.,   {Oya} I.,  2013, \mn@doi
  [\mnras] {10.1093/mnras/sts022}, \href
  {https://ui.adsabs.harvard.edu/abs/2013MNRAS.428..220H} {428, 220}

\bibitem[\protect\citeauthoryear{{Hoffman}}{{Hoffman}}{2019}]{hoffman2019}
{Hoffman} J.,  2019, PhD thesis, Princeton University

\bibitem[\protect\citeauthoryear{{Hosenie}, {Lyon}, {Stappers}  \&
  {Mootoovaloo}}{{Hosenie} et~al.}{2019}]{hosenie2019}
{Hosenie} Z.,  {Lyon} R.~J.,  {Stappers} B.~W.,   {Mootoovaloo} A.,  2019,
  \mn@doi [\mnras] {10.1093/mnras/stz1999}, \href
  {https://ui.adsabs.harvard.edu/abs/2019MNRAS.488.4858H} {488, 4858}

\bibitem[\protect\citeauthoryear{{Hosenie}, {Lyon}, {Stappers}, {Mootoovaloo}
  \& {McBride}}{{Hosenie} et~al.}{2020}]{hosenie2020}
{Hosenie} Z.,  {Lyon} R.,  {Stappers} B.,  {Mootoovaloo} A.,   {McBride} V.,
  2020, \mn@doi [\mnras] {10.1093/mnras/staa642}, \href
  {https://ui.adsabs.harvard.edu/abs/2020MNRAS.493.6050H} {493, 6050}

\bibitem[\protect\citeauthoryear{{Huijse}, {Estevez}, {Protopapas}, {Zegers}
  \& {Principe}}{{Huijse} et~al.}{2012}]{huijse2012}
{Huijse} P.,  {Estevez} P.~A.,  {Protopapas} P.,  {Zegers} P.,   {Principe}
  J.~C.,  2012, \mn@doi [IEEE Transactions on Signal Processing]
  {10.1109/TSP.2012.2204260}, \href
  {https://ui.adsabs.harvard.edu/abs/2012ITSP...60.5135H} {60, 5135}

\bibitem[\protect\citeauthoryear{{Ivezic} et~al.,}{{Ivezic}
  et~al.}{2008}]{lsst}
{Ivezic} Z.,  et~al., 2008, preprint, \href
  {http://adsabs.harvard.edu/abs/2008arXiv0805.2366I} {} (\mn@eprint {arXiv}
  {0805.2366})

\bibitem[\protect\citeauthoryear{{Ivezi{\'c}} et~al.,}{{Ivezi{\'c}}
  et~al.}{2019}]{lsst-2019}
{Ivezi{\'c}} {\v{Z}}.,  et~al., 2019, \mn@doi [\apj]
  {10.3847/1538-4357/ab042c}, \href
  {https://ui.adsabs.harvard.edu/abs/2019ApJ...873..111I} {873, 111}

\bibitem[\protect\citeauthoryear{{Jayasinghe} et~al.,}{{Jayasinghe}
  et~al.}{2018}]{jayasinghe2018}
{Jayasinghe} T.,  et~al., 2018, \mnras, \href
  {https://ui.adsabs.harvard.edu/#abs/2018MNRAS.477.3145J} {477, 3145}

\bibitem[\protect\citeauthoryear{{Jayasinghe} et~al.,}{{Jayasinghe}
  et~al.}{2019}]{jayasinghe2019}
{Jayasinghe} T.,  et~al., 2019, \mn@doi [\mnras] {10.1093/mnras/stz844}, \href
  {https://ui.adsabs.harvard.edu/abs/2019MNRAS.486.1907J} {486, 1907}

\bibitem[\protect\citeauthoryear{{Jimenez} \& {Landgrebe}}{{Jimenez} \&
  {Landgrebe}}{1998}]{jimenez1998}
{Jimenez} L.~O.,  {Landgrebe} D.~A.,  1998, \mn@doi [{IEEE Transactions on
  Systems, Man, and Cybernetics, Part C (Applications and Reviews)}]
  {10.1109/5326.661089}, 28, 39

\bibitem[\protect\citeauthoryear{{Kim} \& {Bailer-Jones}}{{Kim} \&
  {Bailer-Jones}}{2016}]{kim2016}
{Kim} D.-W.,  {Bailer-Jones} C. A.~L.,  2016, \mn@doi [\aap]
  {10.1051/0004-6361/201527188}, \href
  {https://ui.adsabs.harvard.edu/abs/2016A&A...587A..18K} {587, A18}

\bibitem[\protect\citeauthoryear{{Kim} \& {White}}{{Kim} \&
  {White}}{2004}]{kim2004}
{Kim} T.,  {White} H.,  2004, \mn@doi [Finance Research Letters]
  {https://doi.org/10.1016/S1544-6123(03)00003-5}, 1, 56

\bibitem[\protect\citeauthoryear{{Kim}, {Protopapas}, {Byun}, {Alcock},
  {Khardon}  \& {Trichas}}{{Kim} et~al.}{2011}]{kim2011}
{Kim} D.-W.,  {Protopapas} P.,  {Byun} Y.-I.,  {Alcock} C.,  {Khardon} R.,
  {Trichas} M.,  2011, \mn@doi [\apj] {10.1088/0004-637X/735/2/68}, \href
  {https://ui.adsabs.harvard.edu/abs/2011ApJ...735...68K} {735, 68}

\bibitem[\protect\citeauthoryear{{Kim}, {Protopapas}, {Bailer-Jones}, {Byun},
  {Chang}, {Marquette}  \& {Shin}}{{Kim} et~al.}{2014}]{kim2014}
{Kim} D.-W.,  {Protopapas} P.,  {Bailer-Jones} C. A.~L.,  {Byun} Y.-I.,
  {Chang} S.-W.,  {Marquette} J.-B.,   {Shin} M.-S.,  2014, \mn@doi [\aap]
  {10.1051/0004-6361/201323252}, \href
  {https://ui.adsabs.harvard.edu/abs/2014A&A...566A..43K} {566, A43}

\bibitem[\protect\citeauthoryear{{Kohonen}}{{Kohonen}}{1990}]{som}
{Kohonen} T.,  1990, \mn@doi [Proceedings of the IEEE] {10.1109/5.58325}, 78,
  1464

\bibitem[\protect\citeauthoryear{{Kuminski}, {George}, {Wallin}  \&
  {Shamir}}{{Kuminski} et~al.}{2014}]{kuminski2014}
{Kuminski} E.,  {George} J.,  {Wallin} J.,   {Shamir} L.,  2014, \mn@doi
  [\pasp] {10.1086/678977}, \href
  {https://ui.adsabs.harvard.edu/abs/2014PASP..126..959K} {126, 959}

\bibitem[\protect\citeauthoryear{{Larson}, {Beshore}, {Hill}, {Christensen},
  {McLean}, {Kolar}, {McNaught}  \& {Garradd}}{{Larson}
  et~al.}{2003}]{larson2003}
{Larson} S.,  {Beshore} E.,  {Hill} R.,  {Christensen} E.,  {McLean} D.,
  {Kolar} S.,  {McNaught} R.,   {Garradd} G.,  2003, in AAS/Division for
  Planetary Sciences Meeting Abstracts \#35. p. 36.04

\bibitem[\protect\citeauthoryear{{Liu}, {Ting}  \& {Zhou}}{{Liu}
  et~al.}{2012}]{liu2012}
{Liu} F.~T.,  {Ting} K.~M.,   {Zhou} Z.-H.,  2012, \mn@doi [ACM Trans. Knowl.
  Discov. Data] {10.1145/2133360.2133363}, 6, 39

\bibitem[\protect\citeauthoryear{{Lloyd}}{{Lloyd}}{1982}]{kmeans}
{Lloyd} S.,  1982, \mn@doi [IEEE Transactions on Information Theory]
  {10.1109/TIT.1982.1056489}, 28, 129

\bibitem[\protect\citeauthoryear{{Lomb}}{{Lomb}}{1976}]{periodogram1}
{Lomb} N.~R.,  1976, \mn@doi [\apss] {10.1007/BF00648343}, \href
  {https://ui.adsabs.harvard.edu/abs/1976Ap&SS..39..447L} {39, 447}

\bibitem[\protect\citeauthoryear{{Mackenzie}, {Pichara}  \&
  {Protopapas}}{{Mackenzie} et~al.}{2016}]{mackenzie2016}
{Mackenzie} C.,  {Pichara} K.,   {Protopapas} P.,  2016, \mn@doi [\apj]
  {10.3847/0004-637X/820/2/138}, \href
  {https://ui.adsabs.harvard.edu/abs/2016ApJ...820..138M} {820, 138}

\bibitem[\protect\citeauthoryear{{Marrese}, {Marinoni}, {Fabrizio}  \&
  {Altavilla}}{{Marrese} et~al.}{2019}]{marrese2019}
{Marrese} P.~M.,  {Marinoni} S.,  {Fabrizio} M.,   {Altavilla} G.,  2019,
  \mn@doi [\aap] {10.1051/0004-6361/201834142}, \href
  {https://ui.adsabs.harvard.edu/abs/2019A&A...621A.144M} {621, A144}

\bibitem[\protect\citeauthoryear{{McInnes}, {Healy}  \& {Astels}}{{McInnes}
  et~al.}{2017}]{hdbscan}
{McInnes} L.,  {Healy} J.,   {Astels} S.,  2017, \mn@doi [The Journal of Open
  Source Software] {10.21105/joss.00205}, \href
  {https://ui.adsabs.harvard.edu/abs/2017JOSS....2..205M} {2, 205}

\bibitem[\protect\citeauthoryear{{McInnes}, {Healy}, {Saul}  \&
  {Gro{\ss}berger}}{{McInnes} et~al.}{2018}]{umap}
{McInnes} L.,  {Healy} J.,  {Saul} N.,   {Gro{\ss}berger} L.,  2018, \mn@doi
  [Journal of Open Source Software] {10.21105/joss.00861}, 3, 861

\bibitem[\protect\citeauthoryear{{Minniti} et~al.,}{{Minniti}
  et~al.}{2010}]{vvvsurvey}
{Minniti} D.,  et~al., 2010, \mn@doi [\na] {10.1016/j.newast.2009.12.002},
  \href {https://ui.adsabs.harvard.edu/abs/2010NewA...15..433M} {15, 433}

\bibitem[\protect\citeauthoryear{{Molnar}, {Sanders}, {Smith}, {Belokurov},
  {Lucas}  \& {Minniti}}{{Molnar} et~al.}{2022}]{molnar2022}
{Molnar} T.~A.,  {Sanders} J.~L.,  {Smith} L.~C.,  {Belokurov} V.,  {Lucas} P.,
    {Minniti} D.,  2022, \mn@doi [\mnras] {10.1093/mnras/stab3116}, \href
  {https://ui.adsabs.harvard.edu/abs/2022MNRAS.509.2566M} {509, 2566}

\bibitem[\protect\citeauthoryear{{Mowlavi}}{{Mowlavi}}{2014}]{mowlavi2014}
{Mowlavi} N.,  2014, \mn@doi [\aap] {10.1051/0004-6361/201322648}, \href
  {https://ui.adsabs.harvard.edu/abs/2014A&A...568A..78M} {568, A78}

\bibitem[\protect\citeauthoryear{{Naul}, {van der Walt}, {Crellin-Quick},
  {Bloom}  \& {P{\'e}rez}}{{Naul} et~al.}{2016}]{naul2016}
{Naul} B.,  {van der Walt} S.,  {Crellin-Quick} A.,  {Bloom} J.~S.,
  {P{\'e}rez} F.,  2016, arXiv e-prints, \href
  {https://ui.adsabs.harvard.edu/abs/2016arXiv160904504N} {p. arXiv:1609.04504}

\bibitem[\protect\citeauthoryear{{Nun}, {Protopapas}, {Sim}, {Zhu}, {Dave},
  {Castro}  \& {Pichara}}{{Nun} et~al.}{2015}]{nun2015}
{Nun} I.,  {Protopapas} P.,  {Sim} B.,  {Zhu} M.,  {Dave} R.,  {Castro} N.,
  {Pichara} K.,  2015, arXiv e-prints, \href
  {https://ui.adsabs.harvard.edu/abs/2015arXiv150600010N} {p. arXiv:1506.00010}

\bibitem[\protect\citeauthoryear{{Paatero} \& {Tapper}}{{Paatero} \&
  {Tapper}}{1994}]{patero1994}
{Paatero} P.,  {Tapper} U.,  1994, \mn@doi [Environmetrics]
  {https://doi.org/10.1002/env.3170050203}, 5, 111

\bibitem[\protect\citeauthoryear{{Pawlak} et~al.,}{{Pawlak}
  et~al.}{2013}]{pawlak2013}
{Pawlak} M.,  et~al., 2013, \actaa, \href
  {https://ui.adsabs.harvard.edu/abs/2013AcA....63..323P} {63, 323}

\bibitem[\protect\citeauthoryear{{Pearson}}{{Pearson}}{1901}]{pearson1901}
{Pearson} K.,  1901, \mn@doi [The London, Edinburgh, and Dublin Philosophical
  Magazine and Journal of Science] {10.1080/14786440109462720}, 2, 559

\bibitem[\protect\citeauthoryear{{P{\'e}rez-Ortiz}, {Garc{\'\i}a-Varela},
  {Quiroz}, {Sabogal}  \& {Hern{\'a}ndez}}{{P{\'e}rez-Ortiz}
  et~al.}{2017}]{perez2017}
{P{\'e}rez-Ortiz} M.~F.,  {Garc{\'\i}a-Varela} A.,  {Quiroz} A.~J.,  {Sabogal}
  B.~E.,   {Hern{\'a}ndez} J.,  2017, \aap, \href
  {https://ui.adsabs.harvard.edu/#abs/2017A&A...605A.123P} {605, A123}

\bibitem[\protect\citeauthoryear{{Pichara} \& {Protopapas}}{{Pichara} \&
  {Protopapas}}{2013}]{pichara2013}
{Pichara} K.,  {Protopapas} P.,  2013, \mn@doi [\apj]
  {10.1088/0004-637X/777/2/83}, \href
  {https://ui.adsabs.harvard.edu/abs/2013ApJ...777...83P} {777, 83}

\bibitem[\protect\citeauthoryear{{Pichara}, {Protopapas}  \&
  {Le{\'o}n}}{{Pichara} et~al.}{2016}]{pichara2016}
{Pichara} K.,  {Protopapas} P.,   {Le{\'o}n} D.,  2016, \mn@doi [\apj]
  {10.3847/0004-637X/819/1/18}, \href
  {https://ui.adsabs.harvard.edu/abs/2016ApJ...819...18P} {819, 18}

\bibitem[\protect\citeauthoryear{{Pietrukowicz} et~al.,}{{Pietrukowicz}
  et~al.}{2013}]{pietru2013}
{Pietrukowicz} P.,  et~al., 2013, \actaa, \href
  {https://ui.adsabs.harvard.edu/abs/2013AcA....63..115P} {63, 115}

\bibitem[\protect\citeauthoryear{{Pojmanski}}{{Pojmanski}}{2002}]{asas}
{Pojmanski} G.,  2002, \actaa, \href
  {https://ui.adsabs.harvard.edu/abs/2002AcA....52..397P} {52, 397}

\bibitem[\protect\citeauthoryear{{Pollacco} et~al.,}{{Pollacco}
  et~al.}{2006}]{superwasp}
{Pollacco} D.~L.,  et~al., 2006, \mn@doi [\pasp] {10.1086/508556}, \href
  {https://ui.adsabs.harvard.edu/abs/2006PASP..118.1407P} {118, 1407}

\bibitem[\protect\citeauthoryear{{Press} \& {Rybicki}}{{Press} \&
  {Rybicki}}{1989}]{periodogram3}
{Press} W.~H.,  {Rybicki} G.~B.,  1989, \mn@doi [\apj] {10.1086/167197}, \href
  {https://ui.adsabs.harvard.edu/abs/1989ApJ...338..277P} {338, 277}

\bibitem[\protect\citeauthoryear{{Richards} et~al.,}{{Richards}
  et~al.}{2011}]{richards2011}
{Richards} J.~W.,  et~al., 2011, \mn@doi [\apj] {10.1088/0004-637X/733/1/10},
  \href {https://ui.adsabs.harvard.edu/abs/2011ApJ...733...10R} {733, 10}

\bibitem[\protect\citeauthoryear{{Rimoldini} et~al.,}{{Rimoldini}
  et~al.}{2019}]{rimoldini2019}
{Rimoldini} L.,  et~al., 2019, \mn@doi [\aap] {10.1051/0004-6361/201834616},
  \href {https://ui.adsabs.harvard.edu/abs/2019A&A...625A..97R} {625, A97}

\bibitem[\protect\citeauthoryear{{Samus'}, {Kazarovets}, {Durlevich}, {Kireeva}
   \& {Pastukhova}}{{Samus'} et~al.}{2017}]{samus2017}
{Samus'} N.~N.,  {Kazarovets} E.~V.,  {Durlevich} O.~V.,  {Kireeva} N.~N.,
  {Pastukhova} E.~N.,  2017, \mn@doi [Astronomy Reports]
  {10.1134/S1063772917010085}, \href
  {https://ui.adsabs.harvard.edu/abs/2017ARep...61...80S} {61, 80}

\bibitem[\protect\citeauthoryear{{Saxena}, {Prasad}, {Gupta}, {Bharill},
  {Patel}, {Tiwari}, {Er}  \& {Lin}}{{Saxena} et~al.}{2017}]{saxena2017}
{Saxena} A.,  {Prasad} M.,  {Gupta} A.,  {Bharill} N.,  {Patel} O.,  {Tiwari}
  A.,  {Er} M.,   {Lin} C.,  2017, \mn@doi [Neurocomputing]
  {10.1016/j.neucom.2017.06.053}, 267, 664

\bibitem[\protect\citeauthoryear{{Scargle}}{{Scargle}}{1982}]{periodogram2}
{Scargle} J.~D.,  1982, \mn@doi [\apj] {10.1086/160554}, \href
  {https://ui.adsabs.harvard.edu/abs/1982ApJ...263..835S} {263, 835}

\bibitem[\protect\citeauthoryear{{Shapley}}{{Shapley}}{1914}]{shapley1914}
{Shapley} H.,  1914, \mn@doi [\apj] {10.1086/142137}, \href
  {https://ui.adsabs.harvard.edu/abs/1914ApJ....40..448S} {40, 448}

\bibitem[\protect\citeauthoryear{{Soszynski} et~al.,}{{Soszynski}
  et~al.}{2008}]{ogle3}
{Soszynski} I.,  et~al., 2008, \actaa, \href
  {https://ui.adsabs.harvard.edu/abs/2008AcA....58..163S} {58, 163}

\bibitem[\protect\citeauthoryear{{Soszy{\'n}ski}, {Wood}  \&
  {Udalski}}{{Soszy{\'n}ski} et~al.}{2013}]{sosz-2013}
{Soszy{\'n}ski} I.,  {Wood} P.~R.,   {Udalski} A.,  2013, \mn@doi [\apj]
  {10.1088/0004-637X/779/2/167}, \href
  {https://ui.adsabs.harvard.edu/abs/2013ApJ...779..167S} {779, 167}

\bibitem[\protect\citeauthoryear{{Soszy{\'n}ski} et~al.,}{{Soszy{\'n}ski}
  et~al.}{2015}]{sozy2015}
{Soszy{\'n}ski} I.,  et~al., 2015, \memsai, \href
  {https://ui.adsabs.harvard.edu/abs/2015MmSAI..86..257S} {86, 257}

\bibitem[\protect\citeauthoryear{{Stetson}}{{Stetson}}{1996}]{stetson1996}
{Stetson} P.~B.,  1996, \mn@doi [\pasp] {10.1086/133808}, \href
  {https://ui.adsabs.harvard.edu/abs/1996PASP..108..851S} {108, 851}

\bibitem[\protect\citeauthoryear{{Szubert}, {Cole}, {Monaco}  \&
  {Drozdov}}{{Szubert} et~al.}{2019}]{ivis}
{Szubert} B.,  {Cole} J.~E.,  {Monaco} C.,   {Drozdov} I.,  2019, \mn@doi
  [Scientific Reports] {10.1038/s41598-019-45301-0}, \href
  {https://ui.adsabs.harvard.edu/abs/2019NatSR...9.8914S} {9, 8914}

\bibitem[\protect\citeauthoryear{{Tenenbaum}, {de Silva}  \&
  {Langford}}{{Tenenbaum} et~al.}{2000}]{isomap}
{Tenenbaum} J.~B.,  {de Silva} V.,   {Langford} J.~C.,  2000, \mn@doi [Science]
  {10.1126/science.290.5500.2319}, \href
  {https://ui.adsabs.harvard.edu/abs/2000Sci...290.2319T} {290, 2319}

\bibitem[\protect\citeauthoryear{{Valenzuela} \& {Pichara}}{{Valenzuela} \&
  {Pichara}}{2018}]{valenzuela2018}
{Valenzuela} L.,  {Pichara} K.,  2018, \mn@doi [\mnras]
  {10.1093/mnras/stx2913}, \href
  {https://ui.adsabs.harvard.edu/abs/2018MNRAS.474.3259V} {474, 3259}

\bibitem[\protect\citeauthoryear{{\noopsort{VanderMaaten}}{van der Maaten L.,
  Hinton, G.}}{{\noopsort{VanderMaaten}}{van der Maaten L., Hinton,
  G.}}{2008}]{tsne}
{\noopsort{VanderMaaten}}{van der Maaten L., Hinton, G.} 2008, Journal of
  Machine Learning Research, 9, 2579

\bibitem[\protect\citeauthoryear{{\noopsort{Vanengelen}}{van Engelen, J.~E.,
  Hoos, H.~H.}}{{\noopsort{Vanengelen}}{van Engelen, J.~E., Hoos,
  H.~H.}}{2020}]{engelen2020}
{\noopsort{Vanengelen}}{van Engelen, J.~E., Hoos, H.~H.} 2020, \mn@doi [Machine
  Learning] {10.1007/s10994-019-05855-6}, 109, 373

\bibitem[\protect\citeauthoryear{{Virtanen} et~al.,}{{Virtanen}
  et~al.}{2020}]{scipy}
{Virtanen} P.,  et~al., 2020, \mn@doi [Nature Methods]
  {10.1038/s41592-019-0686-2}, \href
  {https://ui.adsabs.harvard.edu/abs/2020NatMe..17..261V} {17, 261}

\bibitem[\protect\citeauthoryear{{\noopsort{Vonneumann}}{von Neumann,
  J.}}{{\noopsort{Vonneumann}}{von Neumann, J.}}{1941}]{vonneumann1941}
{\noopsort{Vonneumann}}{von Neumann, J.} 1941, \mn@doi [Ann. Math. Statist.]
  {10.1214/aoms/1177731677}, 12, 367

\bibitem[\protect\citeauthoryear{{\noopsort{Vonneumann}}{von Neumann,
  J.}}{{\noopsort{Vonneumann}}{von Neumann, J.}}{1942}]{vonneumann1942}
{\noopsort{Vonneumann}}{von Neumann, J.} 1942, \mn@doi [Ann. Math. Statist.]
  {10.1214/aoms/1177731645}, 13, 86

\bibitem[\protect\citeauthoryear{{Watson}, {Henden}  \& {Price}}{{Watson}
  et~al.}{2006}]{watson2006}
{Watson} C.~L.,  {Henden} A.~A.,   {Price} A.,  2006, Society for Astronomical
  Sciences Annual Symposium, \href
  {https://ui.adsabs.harvard.edu/abs/2006SASS...25...47W} {25, 47}

\bibitem[\protect\citeauthoryear{{Webb} et~al.,}{{Webb}
  et~al.}{2020}]{webb2020}
{Webb} S.,  et~al., 2020, \mn@doi [\mnras] {10.1093/mnras/staa2395}, \href
  {https://ui.adsabs.harvard.edu/abs/2020MNRAS.498.3077W} {498, 3077}

\bibitem[\protect\citeauthoryear{{Wright} et~al.,}{{Wright}
  et~al.}{2010}]{wisemission}
{Wright} E.~L.,  et~al., 2010, \mn@doi [\aj] {10.1088/0004-6256/140/6/1868},
  \href {https://ui.adsabs.harvard.edu/abs/2010AJ....140.1868W} {140, 1868}

\bibitem[\protect\citeauthoryear{{Xu}, {Ho}, {Trac}, {Schneider}, {Poczos}  \&
  {Ntampaka}}{{Xu} et~al.}{2013}]{xu2013}
{Xu} X.,  {Ho} S.,  {Trac} H.,  {Schneider} J.,  {Poczos} B.,   {Ntampaka} M.,
  2013, \mn@doi [\apj] {10.1088/0004-637X/772/2/147}, \href
  {https://ui.adsabs.harvard.edu/abs/2013ApJ...772..147X} {772, 147}

\makeatother
\end{thebibliography}



\appendix

\section{Variable star nomenclature and acronyms} \label{appendix:nomeclature}
In the following list, we define the variable star nomenclature and acronyms used in this work:
\begin{itemize}
    \item RRLYR-RRab: RR Lyrae stars pulsating in the fundamental mode (ab-type RR Lyrae, or RRab).
    \item RRLYR-RRc: RR Lyrae stars pulsating in the first overtone (c-type RR Lyrae, or RRc).
    \item RRLYR-RRd: RR Lyrae stars pulsating in the fundamental and first overtone simultaneously (d-type RR Lyrae, or RRd).
    \item RRLYR-RRe: RR Lyrae stars alledgedly pulsating in the second overtone (e-type RR Lyrae, or RRe).
    \item LPV-OSARG: OSARG-type LPVs, where OSARG stands for OGLE's small-amplitude red giants.
    \item LPV-SRV: LPVs of the semi-regular type (SRVs).
    \item LPV-Mira: Mira-type LPVs.
    \item ECL-EC: Contact eclipsing binary systems. 
    \item ECL-ED: Detached eclipsing binary systems. 
    \item ECL-ESD: Semi-detached eclipsing binary systems.      
    \item ECL-EW/EB: $\beta$ Lyrae (EB) or W UMa (EW) type binaries.
    \item ECL-EA: Algol type binaries.
    \item ECL-PCEB: Post-common envelope binary systems.     
    \item ROT-RSCVn: RS Canum Venaticorum (rotational) variables.
    \item ROT-ELL: Rotating ellipsoidal binary variables. 
    \item CEP-T1$_{\rm F}$: Type I Cepheids pulsating in the fundamental mode.
    \item CEP-T1$_{\rm 1O}$: Type I Cepheids pulsating in the first overtone.
    \item CEP-T1$_{\rm 2O}$: Type I Cepheids pulsating in the second overtone.
    \item CEP-T1$_{\rm M}$: Type I Cepheids pulsating in more than one mode simultaneously.
    \item CEP-T2: Type II Cepheids.    
    \item CEP-A: Anomalous Cepheids.
    \item DSCT-S: $\delta$ Scutis pulsating in a single mode. 
    \item DSCT-M: $\delta$ Scutis pulsating simultaneously in more than one mode.  
    \item DPV: Double periodic variables.      
\end{itemize}

\section{\textsc{umap} algorithm} \label{appendix:umap}
\textsc{umap} is a graph-based non-linear DR algorithm \citep{umap}. It embeds the data into a given number of dimensions in two main phases. The first phase involves obtaining a fuzzy topological representation of the data by building a weighted k-nearest neighbors graph from the high-dimensional data. In the second phase, a low-dimensional layout of this graph is optimized. The embedding is iteratively constructed to be as similar as the high-dimensional graph, and a metric is learned to transform new data to the lower-dimensional embedding.

The \textsc{umap} implementation has several parameters that can be tuned. However, default values produce good results in most cases. Here we list the parameters that were relevant to our experiments:
\begin{itemize}
    \item Number of Components ($n\_components$): The number of dimensions of the constructed embedding. For the visualization, we choose two components. We found that around 20 components are appropriate for retaining most of the data's original properties for clustering and classification.
    \item Number of Neighbors ($n\_neigbors$): The number of neighbors used in constructing the weighted graph. A higher number will capture more of the global structure of the data. In our case, values less than 12 tend to fragment the embedding into many small clusters, and values larger than 30 do not show further radical changes. 
    \item Minimum distance ($min\_dist$): This parameter adjusts the attractive force of the points of the constructed embedding. Smaller values of this parameter create dense structures. This parameter is usually set to 0 to have structures as compact as possible.
    \item Target Value ($target\_value$): This is an optional parameter to use for supervised DR. A value of 0 weighs solely on the data (unsupervised information), and a value of 1 weighs solely on the labels (supervised information). In our experiments, this is set to 0.5 by default.
\end{itemize}

\section{\textsc{hdbscan} algorithm} \label{appendix:hdbscan}
\textsc{hdbscan} is a hierarchical density-based clustering algorithm capable of clustering data and detecting noise in high-density regions of various densities \citep{hdbscan}. It is inspired in the \textsc{dbscan} \citep{ester1996} and \textsc{dbscan*} \citep{campello2013} algorithms. \textsc{hdbscan}  works as follows: First, the data density is estimated using the mutual reachability distance metric. This metric magnifies the spreading of points in sparsity zones and leaves the dense regions intact, making the final clustering more robust to noise. Second, a weighted graph of the data is built with edges representing the data points and the weights being the mutual reachability distances. Then this graph is constrained by adopting its minimum spanning tree and sorting its edges by the mutual reachability distance to construct a hierarchical tree (or dendrogram). Finally, a hierarchy of clusters is constructed from the condensed version of the hierarchical tree. There are two main methods to find clusters in this tree: excess of mass (EOM), which finds the clusters that are more persistent or stable, and the leaf method, which chooses the condensed tree leaves as the clusters.

In summary, to perform clustering with \textsc{hdbscan}, we will usually need to tune only four parameters:
\begin{itemize}
    \item Minimum cluster size ($minPts$): The minimum number of members of a cluster. It is useful when there is an idea of the size of the smallest cluster in the data.
    \item Minimum number of samples ($min\_samples$): The number of samples around a point to be designated as a core point. With this parameter, we are affecting how the weighted graph is constructed. Larger values make the clusters condense into progressively denser regions, implying that more samples will be assigned as noise.
    \item Cluster selection epsilon ($\hat{\epsilon}$): This parameter allows to cluster data with clusters of various sizes and densities, avoiding excessive splitting. Setting small values allows finding clusters at the top of the hierarchy.
    \item Cluster selection method: There are two options available, EOM and leaf methods. EOM is the default \textsc{hdbscan} method to find stable clusters. The leaf method can find nested clusters of different sizes in a large structure of various densities.    
\end{itemize}

\bsp	
\label{lastpage}
\end{document}